\documentclass[10pt,twocolumn,aps,prx,superscriptaddress,longbibliography,eprint]{revtex4-2}

\usepackage{graphicx}
\usepackage{amsmath}
\usepackage{amssymb}
\usepackage{float}
\usepackage{braket}   
\usepackage{enumitem}
\usepackage[dvipsnames]{xcolor}
\usepackage[colorlinks=true, urlcolor=blue, linkcolor=blue,citecolor=blue, hypertexnames=false]{hyperref}
\usepackage[detect-weight=true,separate-uncertainty=true]{siunitx}
\usepackage{comment}
\usepackage[normalem]{ulem}

\usepackage{cleveref}
\crefname{equation}{Eq.}{Eqs.}
\Crefname{equation}{Equation}{Equations}
\crefname{figure}{Fig.}{Figs.}
\Crefname{figure}{Figure}{Figures}
\crefname{section}{Sec.}{Sects.}
\Crefname{section}{Section}{Sections}
\crefname{table}{Table}{Tables}
\crefname{appendix}{Appendix}{Apps.}
\Crefname{appendix}{Appendix}{Apps.}

\newcommand{\h}[1]{\hat{#1}}
\newcommand{\ha}{\hat{a}}
\newcommand{\had}{\hat{a}^\dagger}

\newcommand{\caL}{\mathcal{L}}

\newcommand{\hH}{\hat{H}}

\begin{document}

\title{Measurement-induced state transitions across the fluxonium qubit landscape}

\author{Alex A. Chapple}
\email{alex.arimoto.chapple@usherbrooke.ca}
\affiliation{Institut Quantique and D\'epartement de Physique, Universit\'e de Sherbrooke, Sherbrooke J1K 2R1 Quebec, Canada}

\author{Boris M. Varbanov}
\affiliation{Institut Quantique and D\'epartement de Physique, Universit\'e de Sherbrooke, Sherbrooke J1K 2R1 Quebec, Canada}

\author{Alexander McDonald}
\affiliation{Institut Quantique and D\'epartement de Physique, Universit\'e de Sherbrooke, Sherbrooke J1K 2R1 Quebec, Canada}

\author{Alexandre Blais}
\affiliation{Institut Quantique and D\'epartement de Physique, Universit\'e de Sherbrooke, Sherbrooke J1K 2R1 Quebec, Canada}

\date{\today}

\begin{abstract}
    Understanding the mechanisms that limit high-fidelity readout in circuit quantum electrodynamics is essential for its optimization. Multi-photon resonances are understood to be a limiting factor, causing population transfer from the computational states to higher-energy states under drive. This effect, known as measurement-induced state transitions, has been extensively studied for the transmon qubit. While this exploration has begun for the fluxonium qubit, a systematic study of this effect is lacking. Here, we bridge this gap by theoretically studying measurement-induced state transitions in the fluxonium qubit over a wide range of parameters, comprising essentially all experimentally explored ranges. We find that lighter fluxoniums are less susceptible to these state transitions when compared to their heavier counterparts. We attribute this effect to the combination of lower density of multi-photon resonances, a smaller requisite coupling for a given dispersive shift, and a more harmonic-like structure of the charge operator. We confirm the validity of our analysis by performing time-dependent readout simulations. Finally, we consider the impact of the superinductor's array modes on measurement-induced state transitions over a large range of parameters.
\end{abstract}

\maketitle

\section{Introduction} 
Superconducting circuits are a promising platform for fault-tolerant quantum computation, with transmon qubits at the center of most current effort. In parallel, significant effort has been devoted to advancing the capabilities of an alternative superconducting circuit: the fluxonium qubit~\cite{Vladimir:2009}. Advantages of this qubit include its large anharmonicity, typically on the order of several gigahertz~\cite{Vladimir:2009, Nguyen:2019, Zhang:2021, Nguyen:2022, Bao:2022}, as well as dephasing and relaxation times on the order of $1$ ms or more \cite{Somoroff:2023, Ding:2023, Wang:2025}. Taking advantage of these features, recent works have demonstrated single- and two-qubit gate fidelity exceeding $99.9\%$~\cite{Rower:2024, Schirk:2025, Ding:2023, Zhang:2024, Lin:2025, li:2025, Weiss:2022}. Fluxonium qubits therefore represent a serious alternative as building blocks for a fault-tolerant superconducting quantum processor~\cite{Nguyen:2022}.

Crucially, despite the remarkable advances in the field over the past two decades, both the transmon's and fluxonium's slowest and lowest-fidelity operation remains the readout \cite{GoogleQuantumAI:2023, bista:2025, Nesterov:2024, Stefanski:2024, stefanski:2024_exp, Bothara:2025, Watanabe:2025, liu:2026}. Theoretical~\cite{Sank:2016, Cohen2023, Shillito2022, Khezri:2023, Dumas2024, chapple:2025, chapple_longitudinal:2025} and experimental~\cite{Sank:2016, Khezri:2023, Zihao_wang:2025, fechant:2025, Dai:2026} works on the transmon have demonstrated that its intrinsic non-linearity can be the limiting factor in readout if not properly mitigated. This effect, known in the literature under the names of measurement-induced state transitions (MIST), ionization, and drive-induced unwanted state transitions (DUST), is due to the presence of multi-photon processes that arise in the presence of strong drives. This phenomenology, which is responsible for the ionization of highly-excited atoms under microwave fields~\cite{Breuer1989}, is also present with the fluxonium. This has been explored theoretically by considering the ideal fluxonium-resonator Hamiltonian~\cite{Nesterov:2024} or by adding non-idealities such as two-level systems~\cite{bista:2025, li:2025} or the superinductor's array modes~\cite{singh:2025}. 

Although the basic mechanism underlying the drive-induced transitions in the fluxonium is similar to that in the transmon, several distinct features make a systematic investigation of MIST in the fluxonium necessary. Notably, the matrix elements of the fluxonium’s charge and flux operators differ significantly from that of the transmon, resulting in nontrivial couplings between the computational states and higher-excited states. By contrast, in the transmon, states are predominantly coupled to the states nearest in energy. Second, while the fluxonium's inductive shunt makes it insensitive to gate charge fluctuations, it also introduces array modes~\cite{Ferguson:2013, Zhu_perturbation_2013, Viola:2015, Sorokanich:2024} which can limit readout performance and cause additional dephasing \cite{singh:2025}. Finally, the fluxonium's parameter space is inherently more complex. Transmons only have one dimensionless parameter $E_J/E_C$, which must be large ($E_J/E_C \gg 1$) to ensure gate charge insensitivity. Not only does the fluxonium have two such parameters $E_J/E_C$, $E_L/E_C$, but by varying their ratio we can obtain qualitatively different fluxoniums with varying properties. 

In this work, we present a thorough investigation of the physics of measurement-induced state transitions of fluxonium qubits, considering a wide range of experimentally-relevant parameters. We begin in~\cref{sec:Recap_ionization} by briefly summarizing the mechanism of MIST in dispersive readout. In~\cref{sec:fluxonium_parameter_scan}, we present the results of a large parameter scan of nearly $2 \times 10^6$ fluxoniums parameters and resonator frequencies, results highlighting that lighter fluxoniums are less prone to measurement-induced transitions than their heavier counterpart. The physical reasons for this observation are presented in~\cref{sec:mechanism_fluxonium_ionization}. This is followed in~\cref{sec:readout} by full-time dynamic simulation of the dispersive readout, results confirming that the static quantity reported in~\cref{sec:fluxonium_parameter_scan} is a good proxy to predict when drive-induced transitions can be expected to be the limiting factor of the readout fidelity. In~\cref{sec:array_modes} we expand on work presented in Ref.~\cite{singh:2025} and consider the impact of array modes of the fluxonium's superinductor on MIST. Here, we consider the impact of array modes for a wide range of circuit parameters in the asymmetrically coupled case, as well as include the first and second array modes in our analysis. We conclude in~\cref{sec:conclusion}.

\begin{figure}
    \centering
    \includegraphics[width=\linewidth]{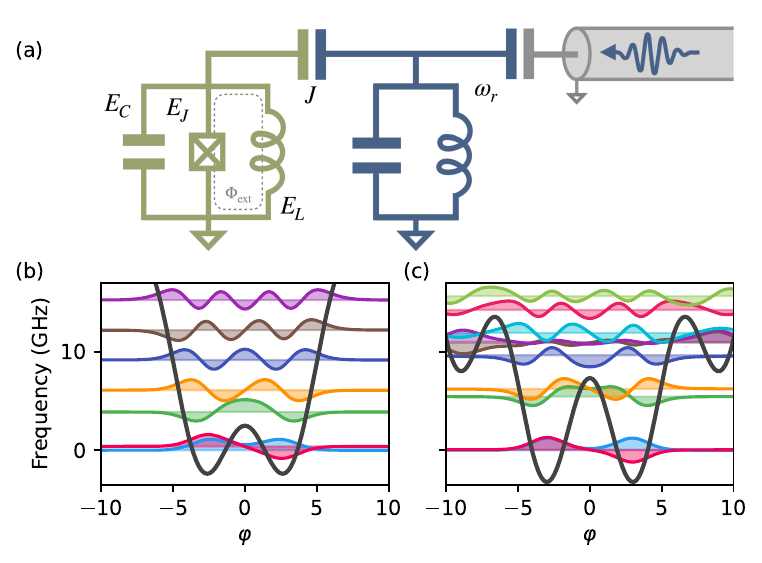}
    \caption{(a) The system we consider is a fluxonium qubit (green) coupled capacitively to a readout resonator (blue). The resonator is coupled to an external feedline (gray). (b) and (c) show the wavefunctions and potential wells for example light and heavy fluxoniums with $E_J / E_C, E_L/E_C = 4, 0.75$ and $E_J / E_C, E_L/E_C = 6, 0.3$ at $\Phi_{\rm{ext}} = 0.5 \Phi_0$, respectively. The wavefunctions were computed using \texttt{scqubits} \cite{groszkowski2021scqubits, chitta2022computer}.
    }
    \label{fig:circuit}
\end{figure}

\section{Dispersive readout and measurement-induced transitions}\label{sec:Recap_ionization}

In this section, we summarize the basics of dispersive readout and its breakdown due to measurement-induced transitions. Our main tool for predicting the occurrence of MIST is the branch analysis~\cite{Boissonneault2010, Shillito2022, Dumas2024} and its semi-classical version, the Floquet branch analysis~\cite{Dumas2024}. For brevity, in this section we only consider the branch analysis. A discussion on the Floquet branch analysis (in the presence of array modes) is presented in~\cref{app:Floquet BA} and we refer the reader to Ref.~\cite{Dumas2024} for a more complete discussion of both methods.

The fluxonium-readout Hamiltonian takes the form ($\hbar$ = 1)
\begin{align}\label{eq:H_full_qubit_rest}
    \hat{H} = \omega_r \had \ha + \hat{H}_{f} - i g (\ha - \had) \hat{n}_f, 
\end{align}
where $\omega_r$ is the resonator's bare frequency, $\ha$ and $\had$ are its lowering and raising operators respectively, $g$ is the coupling strength, and $\hat{H}_f$ is the fluxonium Hamiltonian,
\begin{align}\label{eq:H_f}
    \h{H}_f = 4 E_C \h{n}_f^2 + \frac{E_L}{2}\h{\varphi}_f^2 - E_J \cos{(\h{\varphi}_f - \varphi_{\rm{ext}})},
\end{align}
with $\varphi_{\rm{ext}} = (2\pi / \Phi_0) \Phi_{\rm{ext}}$ and $\Phi_{\rm{ext}}$ the external flux. Here, $\hat{n}_f$ and $\hat{\varphi}_f$ are the fluxonium's charge and phase operator respectively, satisfying the commutation relations $[\hat{\varphi}_f, \hat{n}_f] = i$. In this work we take $\Phi_{\rm{ext}} = 0.5 \Phi_0$. Note that it is also possible to couple the fluxonium and the readout resonator inductively~\cite{Gusenkova:2021, Rieger:2023}. In this case, the coupling takes the form $g(\ha+\ha^\dagger)\hat{\varphi}_f$. In the main text, we focus on the more standard capacitive coupling of~\cref{eq:H_f} and discuss inductive coupling in~\cref{app_sec:inductive_coupling}.

Labeling the fluxonium states by $i_f$ and $j_f$, the condition for dispersive qubit-resonator coupling is given by~\cite{Zhu_perturbation_2013, Nguyen:2022}
\begin{align}\label{eq:dispersive_condition}
    \frac{|g_{i_f,j_f}| \sqrt{n_r+1}}{|\omega_{i_f,j_f} - \omega_r|} \ll 1,
\end{align}
where $g_{i_f,j_f} \equiv g \langle i_f | \hat{n}_f | j_f \rangle$, $n_r$ is the photon number, and $\omega_{i_f,j_f} \equiv \omega_{j_f} - \omega_{i_f}$. This condition ensures that the qubit and resonator do not strongly hybridize. Nevertheless, the coupling is large enough to induce a qubit state-dependent shift $\chi$ on the resonator frequency~\cite{Blais:2004}. By performing a Schrieffer-Wolff transformation $\hat{H}' = e^{\hat{S}}\hat{H} e^{- \hat{S}}$ with the appropriately chosen generator $\hat{S}$ \cite{Blais:2004, Zhu_perturbation_2013}, dropping  off-diagonal terms of order $g^2$ or larger, and projecting onto the qubit subspace, we arrive at the usual dispersive Hamiltonian
\begin{align}\label{eq:H_disperse}
    \hH' \approx (\tilde{\omega}_r + \chi \hat{\sigma}_z)\had \ha + \frac{\tilde{\omega}_q}{2} \hat{\sigma}_z,
\end{align}
where $\tilde{\omega}_r$ and $\tilde{\omega}_q$ are the renormalized resonator and qubit frequency, respectively. In this expression, the dispersive shift $\chi$, which is one of the factors determining the measurement speed, 
is given by
\begin{align}
    \chi
    &=
    \sum_{j \neq 0} \chi_{0j} - \sum_{j \neq 1} \chi_{1j} \nonumber \\
    &= 
    \sum_{j \neq 0}  \frac{\vert g_{0j} \vert^2 \omega_{0j}}{\omega_{0j}^2 - \omega_r^2} 
    - 
    \sum_{j \neq 1} \frac{\vert g_{1j} \vert^2 \omega_{1j}}{\omega_{1j}^2 - \omega_r^2} 
    , \label{eqn:disp_shift}
\end{align}
where we have kept all virtual transitions, including often-neglected counter-rotating terms~\cite{Nguyen:2022}.

Due to multi-photon resonances~\cite{Breuer1989,Sank:2016,Khezri:2023,Shillito2022,Cohen2023,Dumas2024,Xiao2024}, the breakdown of the dispersive Hamiltonian~\cref{eq:H_disperse} can occur at photon numbers that differ drastically from the one predicted by~\cref{eq:dispersive_condition}. There resonances occur when the condition $n\omega_r=m\omega_q(n_r)$ is satisfied, with $n,m$ positive integers and $\omega_q(n_r)$ the ac-Stark-shifted qubit frequency. Because the ac-Stark shift can differ significantly from the perturbative expression~\cref{eq:H_disperse}, the branch analysis is a useful tool to pinpoint the photon numbers at which these resonances occur. 

\begin{figure}
    \centering
    \includegraphics[width=\linewidth]{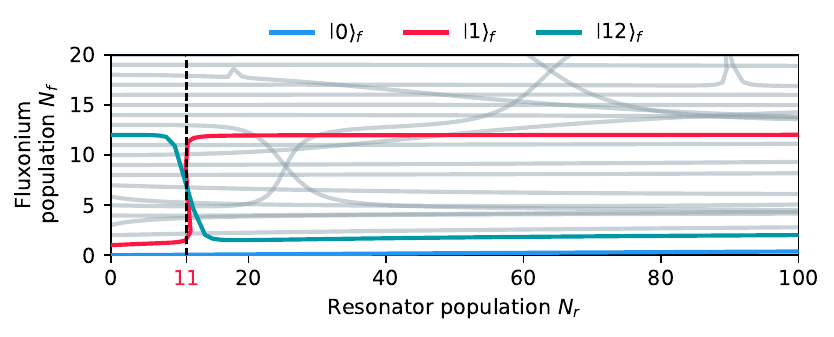}
    \caption{Fluxonium population as a function of the resonator population for the representative light fluxonium ($E_J/E_C = 4.0, E_L/E_C = 0.75$) with $\omega_r /2 \pi = 9.37$ GHz and coupling strength $g / 2\pi = 289$ MHz to the readout resonator. The swapping of the first and twelfth branch indicates a measurement-induced state transition at approximately $11$ photons.}
    \label{fig:branch_analysis_example}
\end{figure}

The starting point of this approach is to numerically obtain the eigenvectors $|\zeta\rangle$ and eigenvalues $E_\zeta$ of the full qubit-resonator Hamiltonian~\cref{eq:H_full_qubit_rest}. Each dressed state $|\overline{i_f,0_r}\rangle$ is then defined as the eigenstate that maximizes the overlap with the corresponding bare zero-photon state $|i_f,0_r\rangle$. All other dressed states are then labeled recursively: Among the remaining unassigned eigenstates $|\zeta\rangle$, the one that maximizes the overlap $|\langle \zeta| \had|\overline{i_f,n_r}\rangle|$ is classified as $|\overline{i_f, n_r+1}\rangle$. The collection of these dressed states for a fixed transmon index $i_f$, constructed by iteratively increasing the photon number, $B_{i_f} = \{|\overline{i_f,n_r}\rangle \: | \: \forall n_r   \: \: \}$ is referred to as a branch. At multi-photon resonances, the composition of the branches involved in the resonance changes abruptly. Thus, examining the branches via the average qubit population $\hat{N}_f \equiv \sum_{i_f} i_f  |i_f\rangle\langle i_f |$ and average photon number $\hat{N}_r \equiv \had \ha$, it is possible to pinpoint at what photon number resonances, and thus ionization or MIST, occurs~\cite{Dumas2024}. An example branch analysis is shown in~\cref{fig:branch_analysis_example} for the representative light fluxonium's parameters shown in~\cref{fig:circuit}(b). An abrupt jump in excitation can be seen for the first excited state when the resonator is populated with approximately $11$ photons.

Following~\cite{Dumas2024}, the critical photon number for MIST in the fluxonium is estimated from the branch analysis in the following way. For a given qubit state $i_f$, we first define $n_{{\rm crit}, i_f}$ as the minimal photon number $\langle \hat{N}_r \rangle$ for which the average fluxonium population reaches $\langle \hat{N}_f \rangle$ = 2 for the ground state and $\langle \hat{N}_f \rangle$ = 3 for the excited state, see the dashed gray vertical lines in \cref{fig:branch_analysis_example}. The critical photon number is then the minimum between the two $n_{\rm crit} = \min_{i_f = 0,1} (n_{{\rm crit}, i_f})$. This criteria for the critical photon number has been shown to agree very well with experimental results~\cite{Dumas2024, fechant:2025, Zihao_wang:2025}.

\section{Fluxonium parameter scan} \label{sec:fluxonium_parameter_scan}

\begin{figure*}
    \centering
    \includegraphics[width=\linewidth]{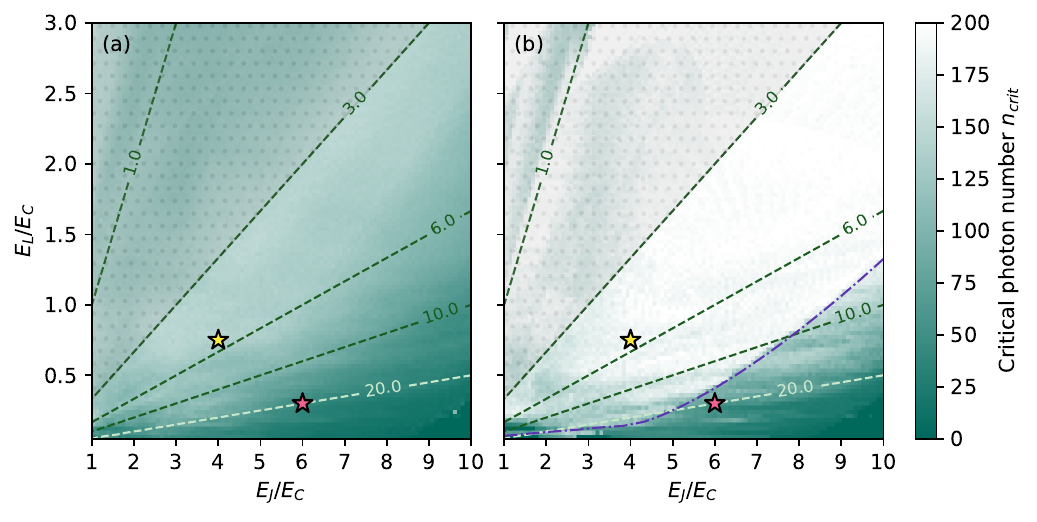}
    \caption{(a) Average critical photon numbers $\bar n_\mathrm{crit}$ across the resonator frequencies $\omega_r / 2\pi \in  [3, 10]$ GHz for each $(E_J/E_C, E_L/E_C)$ combination. We scan over $10^2$ values of $E_J/E_C$ and $E_L/E_C$, as well as $200$ equally spaced resonator frequencies. For each combination $(E_J/E_C, E_L/E_C, \omega_r)$ we find the corresponding coupling strength $g$ that achieves the target dispersive shift of $\chi/2\pi = 2.5$ MHz. Branch analysis is performed for each combination of $(E_J/E_C, E_L/E_C, \omega_r)$ to compute the critical photon numbers, resulting in $2 \times 10^6$ critical photon numbers. The dashed lines indicate $E_J/E_L$ ratios. The yellow (pink) star marks the representative light (heavy) fluxonium's parameter we chose for subsequent analysis, see main text. Shaded-dotted region indicates parameters where the double-well potential does not exist, or is too shallow to be considered as a fluxonium qubit. (b) Maximum of the average critical photon number $\bar n_{\mathrm{crit},\pm\delta}$ across any $2\delta/2\pi = 700$ MHz window of resonator frequencies, for each $(E_J/E_C, E_L/E_C)$ combination. Purple dashed line indicates the contour where $(\omega_{03} - \omega_{12})/2\pi = 1$ GHz.
    }
    \label{fig:giga_scan}
\end{figure*}

Using the branch analysis, we now extract the critical photon numbers $n_{\rm crit}$ of the fluxonium qubit over a large range of parameters. We first discuss the different regimes of the fluxonium qubit, followed by a discussion of the parameter space we explore. We then explain how the critical photon numbers are extracted, and finally discuss the results of our large parameter scan, see~\cref{fig:giga_scan}.

\subsection{Regimes of the fluxonium qubit}

Using the standard analogy of a phase particle of mass set by the capacitance, the first term of~\cref{eq:H_f} plays the role of the kinetic energy while the last two terms are the potential energy of the fluxonium. Varying the dimensionless parameters $E_J/E_C$ and $E_L/E_C$  alters the structure of both the computational states and the higher-excited states. This, in turn, leads to strongly different dispersive shifts---see \cref{eqn:disp_shift}---which play a key role in the onset of measurement-induced transitions. To guide the discussion, it is thus useful to partition the parameter space into distinct regions. 

First, in practice, the charging energy of a fluxonium is more constrained than the Josephson or inductive energies as charging energies greater than $E_C/2\pi = 1$ GHz are difficult to achieve when the fluxonium is capacitively coupled to other qubits, couplers, and readout resonators. To simplify the parameter exploration, we therefore fix the charging energy to $E_C/2\pi = 1~\mathrm{GHz}$ throughout this work. This value is chosen to reflect a typical experimental charging energy used across a broad range of fluxonium implementations \cite{Zhang:2024, Ding:2023, Nguyen:2019, singh_CKA:2025}.

With this choice of $E_C$, the region characterized approximately by $E_J/E_L \in [3, 10]$ corresponds to the ``light" fluxonium regime~\cite{Ding:2023, Nguyen:2022, Nguyen:2019}. Here, the Josephson energy is larger than both the kinetic and harmonic (inductive) energies, leading to the computational subspace being symmetric and anti-symmetric superpositions of persistent current states circulating clockwise or counter-clockwise. Higher states are plasmon-like in nature. An example potential for a light fluxonium with $E_J/E_C = 4$ and $E_L/E_C = 0.75$ is shown in~\cref{fig:circuit}(b). On the other hand, when the ratio $E_J/E_L \gtrsim 10$, the cosine potential plays a larger role and side wells emerge, see~\cref{fig:circuit}(c). Higher excited states of the fluxonium can then become localized in these side wells resulting in fluxon states. Furthermore, the typically higher $E_J/E_C$ ratio results in a larger central potential barrier, lowering the qubit frequency to be of the order of $\omega_q/2\pi \sim 10 - 100$ MHz~\cite{Earnest:2018, Zhang:2021, Zhang:2024}. This is known as the ``heavy'' fluxonium regime. Finally, for $E_J/E_L \lesssim 1$, the harmonic potential dominates the cosine potential, leading to a single well potential. We refer to this as the inductively-shunted transmon regime~\cite{Verney:2019, Hassani:2023, Kalacheva:2024, fasciati:2024, zobrist:2026}. Detailed analysis of this parameter space will be explored in a separate work~\cite{coulombe:2026}.

\subsection{Fluxonium and resonator parameter range} \label{sec:extracting_ncrits}

\Cref{fig:giga_scan} presents the critical photon number obtained from our large parameter scan. With the fixed value of $E_C/2\pi = 1$ GHz, this scan covers the range of Josephson to charging energies $1 \leq E_J/E_C \leq 10$ and ratio of inductive to charging energies $0.05 \leq E_L/E_C \leq 3$. We take $10^2$ equally-spaced values of each dimensionless parameter, for a total of $10^4$ sets of different fluxonium parameters. Because varying these ratios strongly modifies the fluxonium spectrum, leading to low-lying transitions outside the computational subspace that can span several gigahertz, we consider a comparably wide range of resonator frequencies. We thus take $200$ equally space values of the resonator frequency in the range $3$ GHz $\leq \omega_r/2\pi \leq 10$ GHz. Overall, this results in a total of $2 \times 10^6$ parameters for which we compute the branches, and resulting qubit and resonator populations. Finally, our simulations account for 20 states of the fluxonium and 200 states of the resonator. Our results are a comprehensive study of a wide parameter space, encompassing almost all experimentally explored parameter regimes, which required over $500$ hours of computation time on a NVIDIA GH200 GPU.

For a given set of $(E_J/E_C, E_L/E_C, \omega_r)$, there is one remaining free parameter, the qubit-resonator coupling $g$. Since the measurement rate is approximately set by $\sim \chi n$~\cite{Gambetta2006}, for every parameter set we take a value of $g$ (up to a maximum of $g/2\pi = 500$ MHz) such that the dispersive shift is $\chi /2 \pi = 2.5$ MHz. This value is chosen to strike a balance between fast readout and requiring only a modest coupling to the readout resonator for most fluxonium parameters. To ensure that the dispersive approximation remains valid, we ignore any parameters resulting in $g_{i_f j_f}/|\omega_{i_f j_f} - \omega_r| \geq 0.15$, where $i_f = 0,1$ is a computational state and $j_f \neq 0, 1$ is any other state of the fluxonium. Further details on how the coupling strengths were extracted can be found in~\cref{app_sec:scan_details}. 

\subsection{Critical photon numbers and MIST of the fluxonium} \label{sec:ncrit_and_ionization}

\begin{figure*}
    \centering
    \includegraphics[width=\linewidth]{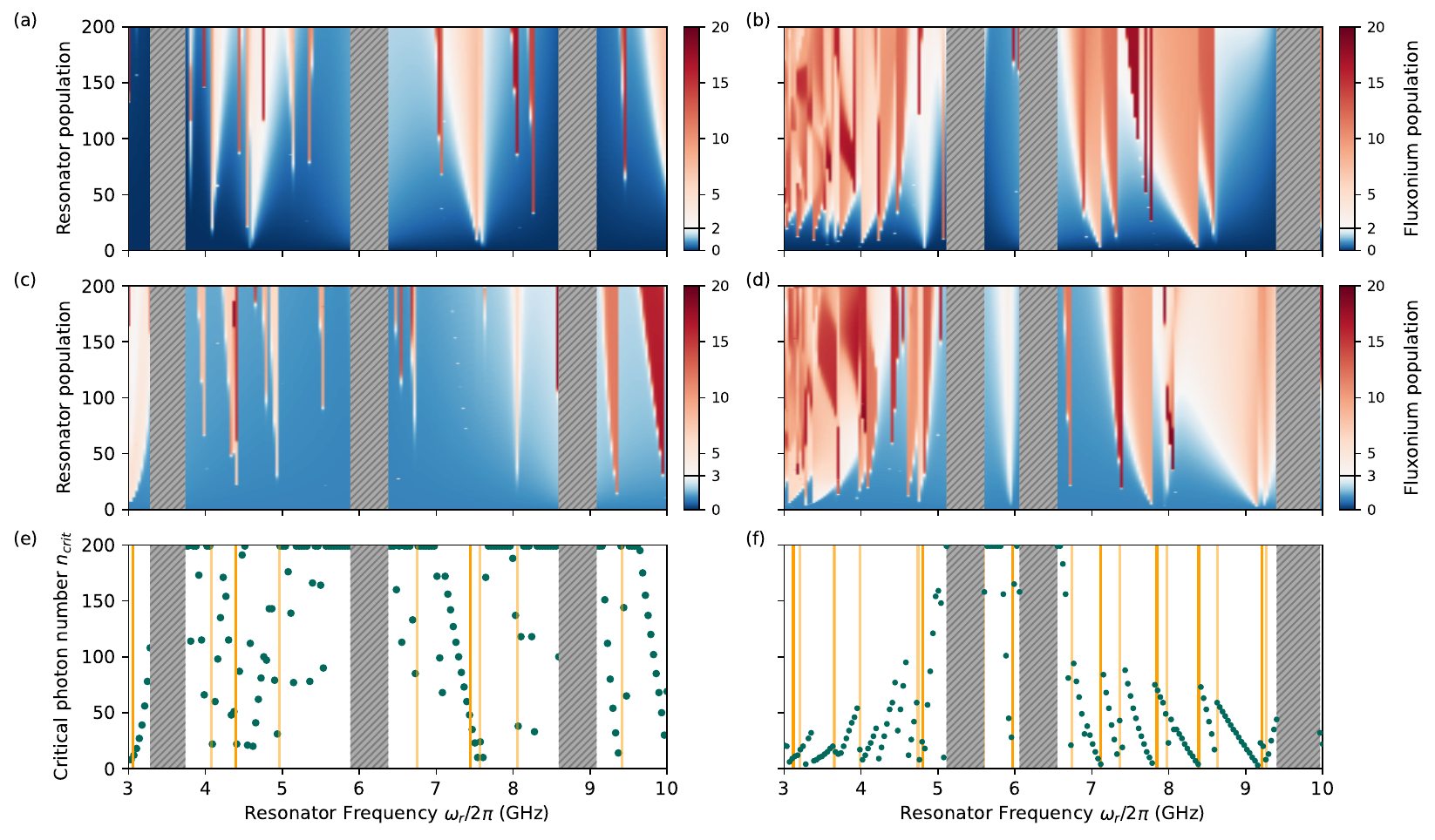}
    \caption{(a, b), (c, d) Fluxonium populations starting in the ground and excited states, respectively, as functions of resonator frequency and population. Panels (a, c) correspond to the light fluxonium, while (b, d) correspond to the heavy fluxonium. (e, f). Critical photon number versus resonator frequency for the light and heavy fluxonium, respectively. Vertical lines mark the predicted locations of 2- (dark orange) and 3-photon (light orange) resonances. The gray regions indicate where the dispersive approximation is invalid.}
    \label{fig:population}
\end{figure*}

\Cref{fig:giga_scan} presents the critical photon number following two approaches. First, in panel (a), for each set of $(E_J/E_C, E_L/E_C)$ the critical photon number is averaged over the full range of resonator frequencies. Because the points at which MIST occurs for a given set $(E_J/E_C, E_L/E_C)$ can be highly sensitive to the choice of resonator frequency, this resonator-frequency averaged critical photon number $\bar n_{\rm crit}$ washes out spurious features, capturing more global properties of the fluxonium defined by that parameter set. As a guide to the eye, the dashed lines in panel (a) correspond to fixed $E_J/E_L$, whose value controls the number of wells in the potential. The shaded-dotted area corresponds to parameters for which the potential does not display a double-well structure, or is too shallow to be considered a fluxonium qubit~\cite{coulombe:2026}. In this large parameter scan, we find a region of large average critical photon numbers approximately in the range $(E_J/E_L,E_J/E_C) \in ([3,6],[2,10])$ corresponding to light fluxonium. We also observe a general trend of decreasing critical photon number when moving towards the heavy regime with $E_J/E_L \gtrsim 10$ and $E_J/E_C \gtrsim 5$. We stress that this does not imply that heavier fluxoniums are necessarily more limited by measurement-induced transitions than lighter ones, but rather that heavier fluxonium tend to exhibit more resonances leading to MIST over the range of resonator frequency range that we considered.

Averaging the critical photon number over the full resonator-frequency range suppresses small, frequency-specific features and allowing to see general trends. However, in practice, the critical photon number need not be large for all $\omega_r$: it is sufficient that it remain large over a reasonably wide range of resonator frequencies over which one or several qubits can be parked for readout. To obtain this more fine-grained information, we first average the critical photon number over a $2\delta/2\pi =700$~MHz-wide window centered on each value of $\omega_r$. We then return the maximum window-averaged critical photon number over the full range of $\omega_r$, $\bar n_{\rm crit,\pm\delta}$. The choice of frequency window $2\delta$ is motivated by large-scale processors, where multiple fluxonium qubits would be measured simultaneously using multiplexed readout. The resonator frequencies would be placed within the bandwidth of the amplifier, with $700$ MHz an  experimentally-relevant range~\cite{Esposito:2021, Evan:2014, Heinsoo:2018, Krinner:2022, Spring:2025}.

In~\cref{fig:giga_scan}(b), we report $\bar n_{\rm crit,\pm\delta}$ for each fluxonium parameters. Here, a large value indicates that there exists some $700$ MHz window in which the critical photon number is on average large. As in~\cref{fig:giga_scan}(a), light fluxoniums have larger critical photon numbers than their heavier counterparts. Indeed, when restricting the average to a 700 MHz window, a significant increase in critical photon number is observed. This increase spills partly into the heavy fluxonium range, until reaching the contour where $(\omega_{03} - \omega_{12})/2\pi = 1$ GHz (purple dot-dashed line). Beyond that contour, the dispersive approximation fails for some transitions, leading to the observed change in critical photon number. The exact position of this contour is a consequence of the choice of $700$ MHz frequency window in our averaging; narrowing this window would move the contour in the direction of the light regime.

Generally speaking, however, \cref{fig:giga_scan}(a) and (b) indicate that the lighter fluxonium are less susceptible to MIST than the heavy fluxoniums. In what follows, we explain why this is the case.

Before turning to that discussion, it is worth emphasizing that the average over the resonator frequency is performed simply to present our simulation data in 2D plots, and that the specific choice of a 700 MHz bandwidth is only one of many possible choices. Less coarse-grained data is available and can be useful in guiding experimental choices. Given a specific fluxonium with ratios $(E_J/E_C,E_L/E_C)$, the landscape of critical photon numbers as a function of resonator frequency $\omega_r$ and coupling strength $g$ is highly complex. The conclusions and intuition we provide based on the coarse-grained data can thus serve as a guide when designing fluxoniums and their readout. However, exploring the fine-grained data to ensure that a given choice of $(E_J/E_C,E_L/E_C, \omega_r)$ is free of deleterious multi-photon transitions remains important.

\section{Mechanisms of MIST in the fluxonium} \label{sec:mechanism_fluxonium_ionization}

In this section, we outline three key factors governing MIST in the fluxonium and highlight how these lead to different behavior in light versus heavy fluxoniums. Throughout this section, we take as examples a light fluxonium with $E_J/E_C = 4.0$ and $E_L/ E_C = 0.75$ and a heavy fluxonium with $E_J/E_C = 6.0$ and $E_L/E_C = 0.3$. These two configurations are indicated by yellow and red stars, respectively, in~\cref{fig:giga_scan}. These serve as representative examples and the qualitative features we discuss below are similar in other light and heavy fluxoniums.

\subsection{Density of multi-photon transitions}

We now consider the density of multi-photon resonances in the fluxonium's spectrum, first focusing on the light fluxonium. \Cref{fig:population}(a,c) show the average fluxonium population of the (a) ground and (c) first excited branch over the full range of resonator frequencies. Blue indicates no significant change in population while a white to red hue indicates that the qubit has been excited outside the computational subspace due to the presence of resonator photons. The gray regions indicate where the dispersive approximation fail. There are several broad features scattered across the range of resonator frequencies: these correspond to two or three multi-photon resonances as shown by the vertical lines in~\cref{fig:population}(e), where we also show the critical photon numbers (dots). More specifically, a vertical orange line indicates a resonator frequency which satisfies $\omega_{i_f j_f} = n \omega_r$ for computational states with $i_f = 0, 1$, $n = 2, 3$, and $j_f$ indexing states outside the computational subspace in the range $2 \le j_f \le 20$ and whose transitions are allowed by parity. At these resonances, the critical photon number is observed to drop rapidly, to increase again until another resonance occurs. Crucially, there exist several large regions in frequency space where these detrimental transitions can be avoided. In these regions, the apparent plateau of the critical photon number at 200 is an artifact of the finite Hilbert space used in our calculations, and the true critical photon number is likely larger.

\Cref{fig:population}(b) and (d) show the ground and excited branch populations of the heavy fluxonium over the range of resonator frequencies. In stark contrast to the light fluxonium, there are essentially no regions of both very large photon numbers and small leakage to higher states. As shown in~\cref{fig:population}(f), this is due to the greater number of two- and three-photon resonances; for the chosen parameters it is nearly double the number as compared to the lighter fluxonium. The larger density of resonances is simply due to the larger number of states at lower energy; compare~\cref{fig:circuit} (b) and (c). This behavior is generic for heavier fluxoniums, since increasing the Josephson energy $E_J$ pushes the higher excited states to lower energies. 

\begin{figure*}
    \centering
    \includegraphics[width=\linewidth]{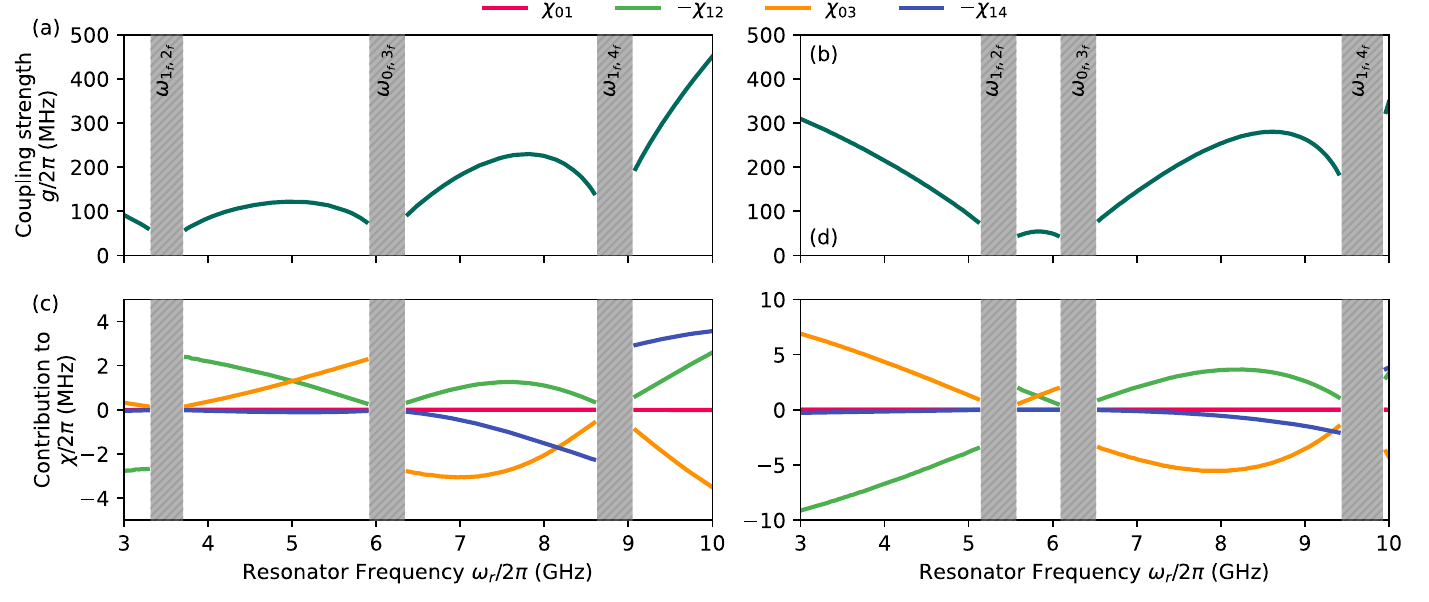}
    \caption{
    (a, b) Required coupling strength $g$ to achieve a target dispersive shift of $\chi/2\pi = 2.5$ MHz for the light and heavy fluxonium, respectively. Grey dashed region indicates where the dispersive approximation is invalid. (c, d) Contributions to the dispersive shift from the first five levels of the representative light and heavy fluxonium, respectively, across resonator frequencies $\omega_r/2\pi = 3$–$10$ GHz. Note that we choose to plot $-\chi_{12}$ and $-\chi_{14}$ since these are the terms contributing to $\chi$, see \cref{eqn:disp_shift}.
    }
    \label{fig:disp_shift_g}
\end{figure*}

\subsection{Average coupling strength} \label{sec:avg_coupling_strength}

In addition to the density of multi-photon transitions, the magnitude of the coupling between the resonator and fluxonium also plays a role in determining the threshold of MIST. As discussed above, in~\cref{fig:giga_scan,fig:population} the dispersive shift  for each fluxonium parameters and resonator frequency $(E_J/E_C, E_L/E_C, \omega_r)$ is fixed to $\chi/2\pi = 2.5$ MHz by choosing the appropriate coupling strength $g$. As shown in~\cref{fig:disp_shift_g}(a,b), this leads to larger coupling strength $g$ for heavy fluxoniums (b) relative to light fluxoniums (a). The reason can be traced back to the structure of the eigenstates and their matrix elements in the charge basis. 

To see this, in~\cref{fig:disp_shift_g}(c,d) we show the contributions to the total dispersive shift from each level of the fluxonium for the light and heavy variant, respectively. Focusing first on the light fluxonium, see~\cref{fig:disp_shift_g}(c),  and on the region between the $\omega_{1_f, 2_f}$ and $\omega_{0_f, 3_f}$ transitions, we see that the dominant contributions to the dispersive shift, namely $\chi_{12}$ and $\chi_{03}$ (see~\cref{eqn:disp_shift}), have the same sign and thus add up. In turn, this leads to a smaller target coupling $g$ to obtain the same $\chi$. The same story can be said of the region between the transitions $\omega_{0_f, 3_f}$ and $\omega_{1_f, 4_f}$. There, the contributions from state $\chi_{03}$ and $\chi_{14}$ add up, while $\chi_{12}$ has an opposite sign and thus destructively interferes.

On the other hand, in the case of the heavy fluxonium, two mechanisms play an important role in setting the required value of $g$ to obtain a target dispersive shift $\chi$. First, in the limit of very large $E_J$, pairs of states, such as the second and third excited states of the fluxonium, become even and odd superpositions of the single-well bound state. This, in turn, implies that $\omega_{0_f, 3_f} \approx \omega_{1_f, 2_f}$, see~\cref{fig:circuit}(c). Therefore, as can be seen in \cref{fig:disp_shift_g}(d), the region where the contributions $\chi_{12}$ and $\chi_{03}$ have the same sign is narrow. Second, the matrix element structure are such that $|\langle 0_f| \hat{n}_f | 3_f \rangle| \approx |\langle 1_f| \hat{n}_f | 2_f \rangle| \gg |\langle 1_f| \hat{n}_f | 4_f \rangle|$. Thus, outside of the frequency range $\omega_{1_f,2_f} \leq \omega_r \leq \omega_{0_f, 3_f}$, $\chi_{12}$ and $\chi_{03}$ still dominate but are nearly equal in magnitude with opposite signs, interfering with one another. This in turn results in a low dispersive shift that can only be compensated by increasing the coupling strength $g$ to the resonator, as seen in~\cref{fig:disp_shift_g}(b).

\begin{figure}
    \centering
    \includegraphics[width=\linewidth]{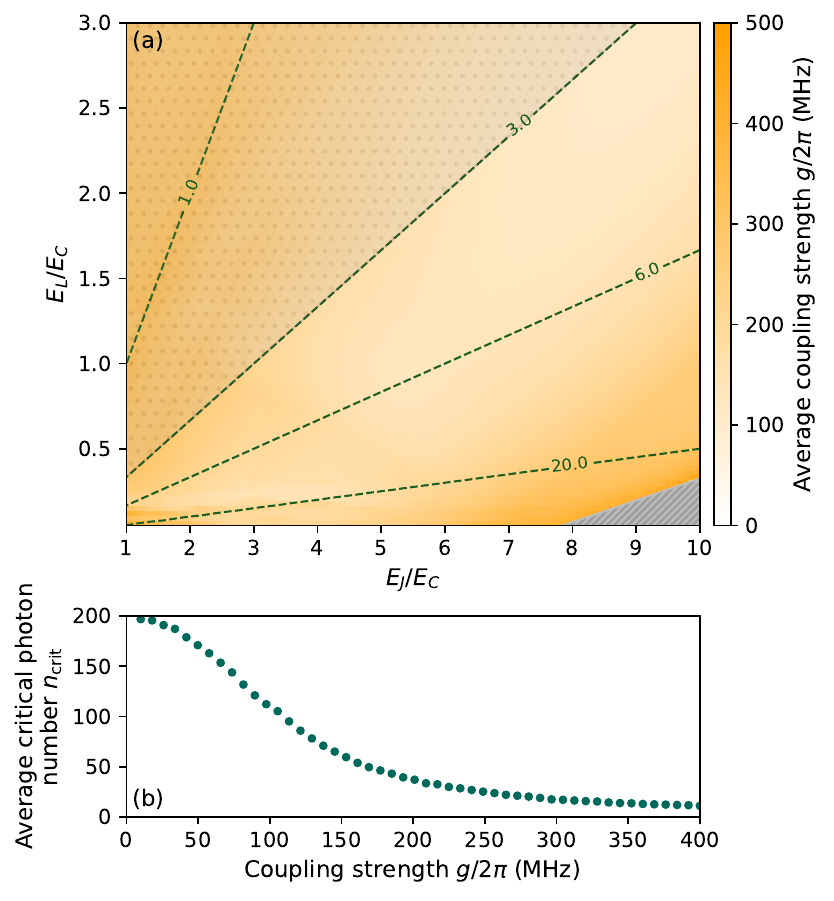}
    \caption{(a) Average coupling strength over resonator frequencies required to achieve the target dispersive shift $\chi / 2\pi = 2.5$ MHz. Dashed lines indicate the $E_J/E_L$ ratio, and the dotted-shaded region on the upper left indicates the parameters where the double-well potential does not exist, or is too shallow to be considered a fluxonium qubit. The shaded region on the bottom right indicates where the average coupling strength exceeds the maximum $g / 2\pi = 500$ MHz. (b) Average critical photon number across resonator frequencies for varying coupling strengths $g$, for the representative heavy fluxonium qubit.
    }
    \label{fig:avg_coupling_strength} 
\end{figure}

To demonstrate that, all other things being equal, a higher coupling facilitates MIST, for each fluxonium $(E_J/E_C, E_L/E_C)$ we plot the average coupling strength $\bar g_{\omega_r}$ required to obtain the target dispersive shift $\chi/2\pi = 2.5$ MHz in~\cref{fig:avg_coupling_strength}. We can confirm that lighter fluxoniums necessitate a smaller coupling than their heavier counterparts. Further, momentarily allowing for a varying dispersive shift, we plot the average critical photon number $\bar n_\mathrm{crit}$ as a function of the coupling $g$ for the chosen representative heavy fluxonium. Unsurprisingly, a larger coupling strength results in a reduction of the critical photon number.

\subsection{Structure of the charge operator}

\begin{figure}
    \centering
    \includegraphics[width=\linewidth]{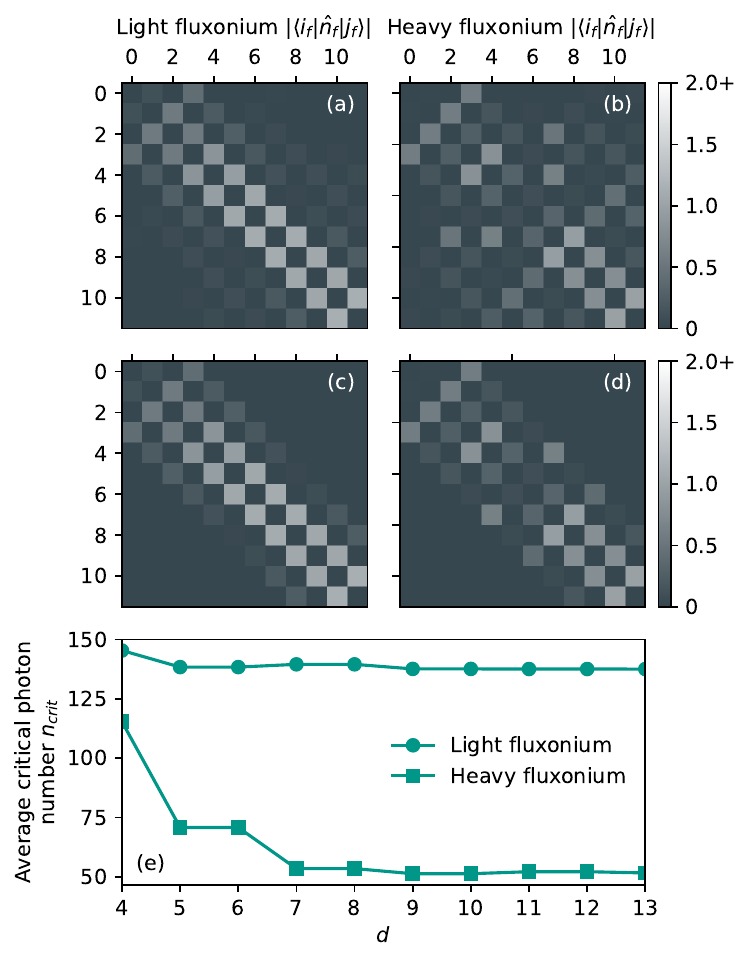}
    \caption{(a), (b) Charge matrix elements of the representative light and heavy fluxoniums, respectively. (c), (d) Charge matrix elements of the representative light and heavy fluxoniums, with only three off-diagonals kept and all other off-diagonal elements set to $0$. (e) Average critical photon number across resonator frequencies ranging between $\omega_r/2\pi = 3 - 10$ GHz when varying $d$ in the modified fluxonium charge operator. See main text and \cref{eqn:charge_matelem_d} for details.
    }
    \label{fig:ncrits_vs_deleted_matelems}
\end{figure}

The structure of the charge operator matrix element is another key factor determining the threshold of measurement-induced transitions. In \cref{fig:ncrits_vs_deleted_matelems} (a) and (b) we plot the charge matrix elements of our representative light and heavy fluxoniums, respectively. While the structure of those matrix elements is mostly dominated by a few diagonal bands in the light case, the matrix elements in the heavy case do not have a simple structure and are instead spread over many transitions. This can be understood as resulting from a competition between the confining quadratic and periodic potential. Indeed, the structure of the higher excited states of the light fluxoniums is largely controlled by the quadratic potential, rendering these states more harmonic-oscillator like in nature and thus leading to charge matrix elements that resemble those of the harmonic oscillator. On the other hand, with their larger $E_J/E_L$ ratio, the eigenstates of heavier fluxoniums are further away from harmonic oscillator states, which breaks selection rules and leads to enhanced charge matrix elements beyond the main off-diagonal.

To probe the impact of charge matrix elements beyond the first few diagonal bands on MIST, we artificially  eliminate them in the charge-charge coupling and using these modified matrices to compute the critical photon numbers. Specifically, for a given positive integer $d \geq 4$, we define
\begin{align}
    \hat{n}_{f}^{(d)}
    =
    \begin{cases}
        \langle i_f | \hat{n}_f | j_f \rangle, &  |i_f-j_f| \leq d  \\
        0, & |i_f-j_f| > d,
    \end{cases}
    \label{eqn:charge_matelem_d}
\end{align}
see~\cref{fig:ncrits_vs_deleted_matelems} (c,d) for an illustrative example. This simplified charge operator is then used in the fluxonium-resonator coupling term of~\cref{eq:H_f} which now reads $-ig (\hat{a} - \hat{a}^\dagger) \hat{n}_f^{(d)}$. Keeping $d \geq 4$ ensures that the dispersive shift remains essentially unchanged in all of the explored parameter space. In~\cref{fig:ncrits_vs_deleted_matelems}(e) we plot the resonator-frequency average critical photon number $\bar n_{\rm crit}$ as a function of $d$. A drastic reduction in the critical photon number is observed for the heavy fluxonium (square symbols) at $d = 5$. This drop results from the addition at $d=5$ of a finite direct coupling between the computational states and the fluxons state $j_f = 5$. There is then another, smaller, decline at $d = 7$ after the critical photon numbers saturates to the value obtained when accounting for the charge operator. In contrast, the critical photon number of the light fluxonium displays only a weak sensitivity on $d$, as the dominant structure of the charge matrix elements is already captured by the few diagonal bands visible in both~\cref{fig:ncrits_vs_deleted_matelems} (a) and (c). This highlights the fact that fluxon states mediate additional detrimental processes. Indeed, because of the stronger non–nearest-neighbor matrix elements compared to that of the light fluxonium, there are more ``paths" from the qubit states to higher excited states, thereby lowering the critical photon numbers across a broad range of parameter space and resonator frequencies in the heavy regime.

\section{Readout simulations} \label{sec:readout}

In the previous section, we obtained the critical photon number using the branch analysis. This method relies on the eigenstates of the undriven fluxonium-resonator Hamiltonian. Here, we compare the critical photon numbers obtained in this way to those obtained from full time-dependent simulations of the dispersive readout including the measurement drive. More precisely, we simulate the readout dynamics using the standard Lindblad master equation
\begin{align}
    \frac{d\hat{\rho}}{dt}= -i[\hat{H} + \h{H}_d,\hat{\rho}] + \kappa \mathcal{D}[\hat{a}]\hat{\rho}, \label{eqn:master_eqn}
\end{align}
where $\h{H}$ is the fluxonium-resonator Hamiltonian of~\cref{eq:H_full_qubit_rest} and $\kappa$ the resonator decay rate. Moreover, $\h{H}_d = -i \epsilon(t) \cos{(\omega_d t)} (\ha - \had)$ is the drive Hamiltonian with $\epsilon(t)$ the drive pulse and $\omega_d$ the drive frequency. In addition to capturing multi-photon resonances, the full time dynamics accounts for the time spent near a given resonance as the photon number changes under the drive, which is important in determining the population transfer and thus whether a MIST event occurs~\cite{Breuer1989,Shillito2022,Dumas2024}. In other words, the full time dynamics provides a consistency check, confirming that the critical photon numbers identified by the branch analysis are indeed expected to lead to a breakdown of QNDness in dispersive readout. Importantly, however, the full time-dependent simulations of the dispersive readout are prohibitively long to scan over large parameter ranges. Indeed, thanks to the relative simplicity of the branch analysis method, it was possible to obtain critical photon numbers for $2\times 10^6$ fluxonium and resonator parameters using GPU acceleration in $\sim 500$ hours on a single Nvidia GH200. Performing the same scan using full-time dynamics would, in contrast, take orders of magnitude longer.

\begin{figure}
    \centering
    \includegraphics[width=\linewidth]{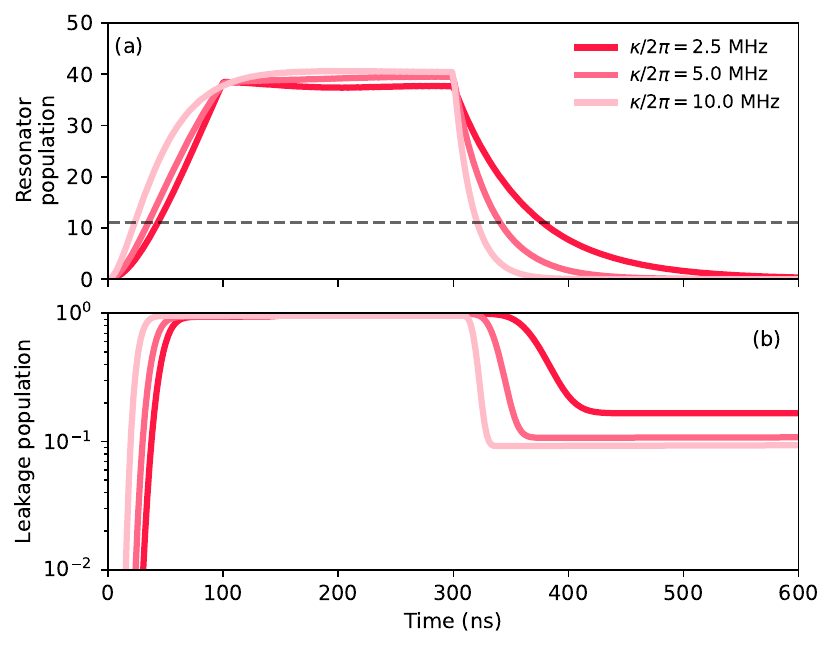}
    \caption{(a) Resonator and (b) leakage population as a function of time during readout when starting in $\vert \overline{1_f, 0_r} \rangle$ for various resonator decay rates $\kappa$. Horizontal dashed line in (a) indicates a resonator population of $11$ photons. The fluxonium, resonator, and coupling parameters are the same as in~\cref{fig:branch_analysis_example}.
    }
    \label{fig:readout_simulation}
\end{figure}

As an illustrative example, we thus focus on the light fluxonium used in the previous section with $(E_J/E_C, E_L/E_C) = (4,0.75)$ and $E_C/2\pi = 1$ GHz. Working with those fluxonium parameters, we consider two sets of resonator frequency and fluxonium-resonator coupling leading to i) low critical photon number and ii) high critical photon number. 

\subsection{Low critical photon numbers}

We first place the resonator frequency at $\omega_r /2 \pi = 9.37$~GHz and take $g / 2\pi = 289$~MHz such that $\chi / 2\pi = 2.5$~MHz. As seen from the branch analysis of~\cref{fig:readout_simulation}(a), this results in a critical photon number $n_{\rm crit} \approx 11$ where the population of the first excited state $i_f = 1$ swaps with $j_f = 12$ (dashed vertical line). 

As discussed in more detail in~\cref{app_sec:readout_simulations}, we employ a standard two-step measurement pulse amplitude $\epsilon(t)$, where the initial step drives the resonator to an average photon number of approximately $\overline{n} \approx 40$ photons in $100$~ns, and the second step aims to hold this population for $200$~ns. Using a two-step pulse allows us to control the rate of change of the photon number, and thus speed at which multi-photon resonances are crossed. We then let the resonator decay freely for $300$~ns, see~\cref{fig:readout_simulation}(b) which shows the average photon population as a function of time. Another parameter that controls the rate of change of the photon number is the resonator decay rate $\kappa$. For this reason, in our simulation we use three different values of this rate: $\kappa/2\pi = 2.5, 5, 10$~MHz.

\begin{figure}
    \centering
    \includegraphics[width=\linewidth]{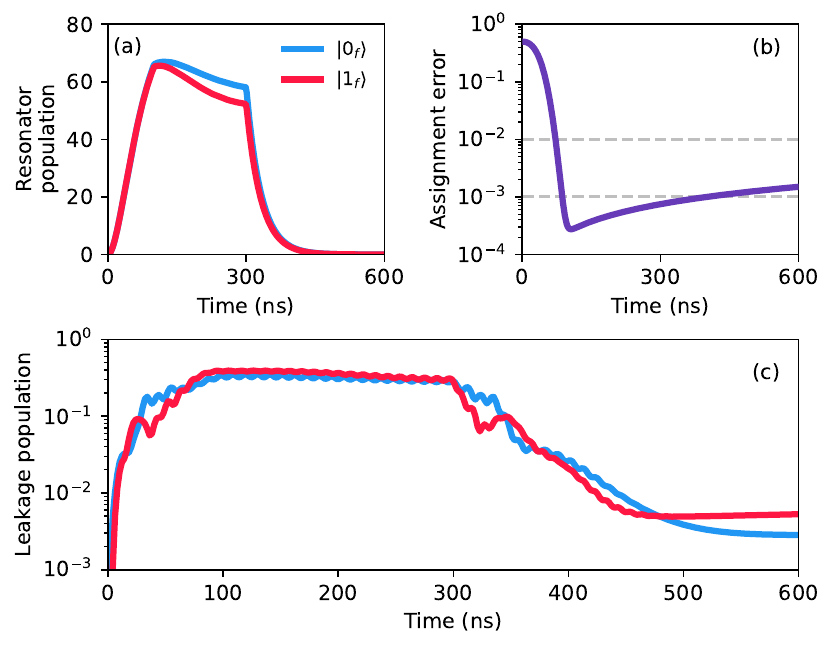}
    \caption{(a) Resonator population, (b) assignment error and (c) leakage population as a function of integration time. Leakage population is defined as $p_l = 1 - (p_{\vert \overline{0}_f \rangle} + p_{\vert \overline{1}_f \rangle})$, where $p_{\vert \overline{i}_f \rangle}$ is the population of the $i$th fluxonium branch defined in \cref{eq:P_i_f_def}. See details in \cref{app_sec:readout_simulations}.
    }
    \label{fig:high_ncrit_readout_sim}
\end{figure}

\Cref{fig:readout_simulation}(c) shows the numerically obtained average leakage population out of the computational subspace as a function of time when initializing the system in $\ket{\overline{1_f,0_r}}$. We define leakage to be any population outside the computational branches, $L(t) = 1-p_{\overline{0}_f}(t) - p_{\overline{1}_f}(t)$, where
\begin{align}\label{eq:P_i_f_def}
    p_{\overline{i}_f}
    \equiv
    \sum_{\overline{n}_r}
    \langle 
    |\overline{i_f, n_r}\rangle \langle \overline{i_f, n_r} |
    \rangle
\end{align}
is the average population of branch $B_{i_f}$. To correctly interpret this quantity it is important to go back to the state identification across a multi-photon transition used in the branch analysis. As a concrete example, consider the first excited state branch (red) in~\cref{fig:branch_analysis_example}(a). At the resonances indicated by the vertical dashed line, the population of this branch swaps with branch $B_{12_f}$ (teal). After this swap, branch $B_{1_f}$ is close in character to bare state $12_f$, while branch $B_{12_f}$ to $1_f$. As a result, there is \emph{no} leakage out of the computational subspace if, at the resonance, the system transitions from branch $B_{1_f}$ to branch $B_{12_f}$. In that case, the readout is QND.

In~\cref{fig:readout_simulation}(b), the leakage $(L)$ is observed to abruptly jump from 0 to nearly unity when the resonator population crosses the critical photon number found from the branch analysis, see the dashed horizontal line in panel (a). This indicates a near perfect transfer from  $B_{1_f}$ to $B_{12_f}$; at this stage this does not correspond to actual leakage, and the readout is still QND. A second jump is observed when the resonator population crosses the critical photon number a second time as the resonator is emptied. The speed of this passage is only controlled by the decay rate $\kappa$ and, being much slower than the rate up, some population remains trapped in $B_{12_f}$. With the resonator now empty, this results in actual leakage, and thus to a non-QND character of the readout. As expected from this discussion, the final leakage increases with decreasing $\kappa$. Furthermore, not only does MIST cause leakage out of the computational subspace, it can also degrade the assignment error. Indeed, if a leaked state has significant overlap in phase space with the ground or excited state, it can be difficult or impossible to distinguish the states with a threshold.

\subsection{High critical photon numbers}

We now place the resonator frequency at $\omega_r /2 \pi = 7.925$~GHz and take $g / 2\pi = 227.5$~MHz, again such that $\chi / 2\pi = 2.5$ MHz. With these parameters, the branch  analysis yields a critical photon number exceeding 200, the size of the resonator Hilbert space used in this analysis, see \cref{fig:high_ncrit_readout_sim_detailed}(a). We use a two-step pulse with a total duration of 300~ns and filling the resonator with approximately 70 photons. We then let the resonator ring down for another 300~ns, see~\cref{fig:high_ncrit_readout_sim}(a) which shows the resonator population with the qubit initialized in the dressed state $\bar 0_f$ (blue) or $\bar 1_f$ (red) with a fixed value of $\kappa/2\pi = 5$~MHz. 

Given that no multi-photon resonances are crossed during this readout, we expect the measurement fidelity to be large. As a confirmation,~\cref{fig:high_ncrit_readout_sim}(b) shows the estimated assignment error $\varepsilon$ assuming a measurement efficiency $\eta = 0.5$ and a qubit lifetime of $T_1 = 200~\mu s$; see~\cref{app_sec:readout_simulations} for more details on readout simulation and assignment error calculation. We find that a readout fidelity of $1-\varepsilon = 99.97\%$ is attainable in an integration time $\tau\sim100$~ns with the chosen parameters. This results is close to the limit $1-\tau/2T_1\sim 99.975\%$ imposed by the finite qubit relaxation time~\cite{Swiadek:2024}.

In addition to the readout fidelity, it is useful to characterize the QND character of the measurement. \Cref{fig:high_ncrit_readout_sim}(c) shows the average leakage population when starting in the ground (blue) or excited (red) state of the fluxonium. We find that, at the end of the readout, $0.29\%$ and $0.55\%$ of the population as leaked out of the computational subspace for the initial ground and excited states, respectively. When starting in the ground state, most of the leakage occurs to states $\vert 3_f \rangle$ and $\vert 5_f \rangle$, whereas starting in the first excited state, leakage predominantly populates $\vert 2_f \rangle$. This behavior is consistent with the large charge matrix elements connecting the computational states to these higher states, see~\cref{fig:ncrits_vs_deleted_matelems}(a), as well as the resonator being placed near the relevant transition frequencies. Leakage reduction units can potentially alleviate the dominant leakage to these states~\cite{Battistel:2021, Varbanov:2020, Marques:2023, Lacroix:2025, Miao:2023}. However, we also observe non-negligible leakage to higher excited states that may be harder to reset with high fidelity. We therefore anticipate that, even when the critical photon number reaches values as high as $200$, leakage into higher excited states -- arising from strong hybridization with states outside the scope of reset protocols -- may constrain the maximum photon number used for readout.

\section{Array modes} \label{sec:array_modes}

\begin{figure}
    \centering
    \includegraphics[width=\linewidth]{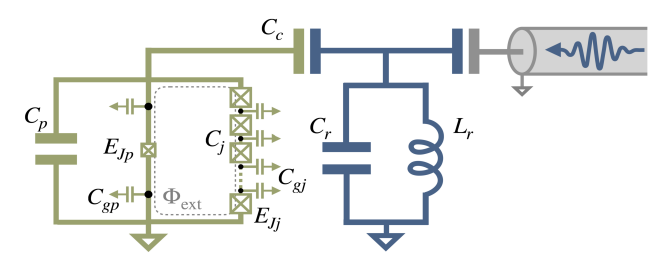}
    \caption{A fluxonium qubit asymmetrically coupled to a readout resonator. Here $C_p$ is the fluxonium capacitance, $C_{gp}$ the capacitance of the phase slip junction to the ground, and $E_{Jp}$ the Josephson energy of the phase slip junction. Moreover, $E_{Jj}$ is the Josephson energy of the array junctions, and $C_{gj}$ is their capacitance to the ground while $C_j$ is the self capacitance of the array junctions. The fluxonium is coupled capacitively to the readout resonator with capacitance $C_c$, and the resonator has a capacitance of $C_r$ and inductance of $L_r$.
    }
    \label{fig:array_circuit}
\end{figure}

We have thus far used the standard fluxonium Hamiltonian~\cref{eq:H_f}. This, however, is an idealized model which ignores additional modes that arise due to the array of Josephson junctions used to realize the fluxonium's quadratic phase potential~\cite{Masluk2012,Ferguson:2013}, see~\cref{fig:array_circuit}. These array modes couple to the qubit degree of freedom via the array junctions' capacitance to ground. Although this coupling to the qubit can be large, the array-modes frequencies are also highly detuned from the qubit's computational states, thus ensuring that~\cref{eq:H_f} remains a good approximation at low energies.

However, under MISTs, the system can access high energy states where such approximations are invalid. The importance of additional modes on measurement-induced transitions has already been explored in different settings~\cite{Zihao_wang:2025,benhayoune:2025,singh:2025, hoyau:2026}. For the fluxonium,~\textcite{singh:2025} recently showed that array modes can lead to drive-induced transitions during readout in situations where they would otherwise be absent, a process they referred to as parametric MIST (PMIST).

The underlying mechanism for these transitions remains multi-photon processes which, given that they are mediated by array modes, depends sensitively on the details of the device. With these findings in mind, in this section we include the presence of array modes in our analysis. Here, we extend the study done in Ref.~\cite{singh:2025} and explore a large parameter space of ground capacitances, resonator frequencies, and dispersive shifts. Moreover, in contrast to Ref.~\cite{singh:2025}, we consider the asymmetric coupling that is most commonly realized in experiments. For the symmetric coupling case, the fluxonium is coupled to both ends of the resonator \cite{Ferguson:2013, Viola:2015, singh:2025}. Here, in contrast, the coupling breaks the inversion symmetry of the junction array as shown in \cref{fig:array_circuit}. Importantly, because of the lack of inversion symmetry, in this case the qubit couples to the lowest array mode which is otherwise decoupled from the qubit in the case of symmetric coupling. In our study, we thus not only account for the second mode but for the two lowest frequency array modes. By accounting for this additional mode, we find that although the qubit–array coupling to the first mode is typically weaker than to the second, its lower frequency can make it the dominant cause of measurement-induced transitions. 

\subsection{Extracting critical photon numbers in the presence of array modes}

The Hamiltonian describing the coupling between the qubit, resonator, and array modes of the circuit of \cref{fig:array_circuit} is given by (see \cref{app_sec:array_modes} for details)
\begin{align}\label{eq:H_f_array}
    \hat{H}
    =
    \omega_r \hat{a}^\dagger \hat{a}
    +
    \hat{H}_f
    -i g_{rf}(\ha-\ha^\dagger) \hat{n}_f
    +
    \hat{H}_{arr}
    +
    \hat{H}_{int}
    .
\end{align}
The first three terms have the same interpretation as in~\cref{eq:H_f} with parameters that depend on the capacitances and inductances of the circuit, see~\cref{app_sec:array_modes} for the full expressions. The penultimate term represents the free array modes Hamiltonian with $\hat{H}_{arr} = \sum_{\mu} \omega_\mu \hat{c}^\dagger_\mu \hat{c}_\mu $, where $\omega_\mu$ is frequency of the $\mu$-th array mode, and $\hat{c}_\mu$ and $\hat{c}^\dagger_\mu$ are its annihilation and creation operators. In writing this expression, we used the standard approximation that the Josephson energy of the array junctions dominate their charging energy.  For an array of $L$ junctions, there are $L-1$ array modes. The last term of~\cref{eq:H_f_array} corresponds to a standard charge-charge coupling between the array modes and the qubit, $g_{f\mu}$, and readout, $g_{r\mu}$, degrees of freedom
\begin{align}
    \hat{H}_{\rm int}
    =
    -
    \sum_{\mu}\left[
    i 
    g_{f \mu}
    \hat{n}_f
    +
    g_{r\mu}
    (\hat{a}-\hat{a}^\dagger)
    \right]
    (\hat{c}_\mu - \hat{c}_\mu^\dagger),
\end{align}
see \cref{app_sec:array_modes} for the expressions for the couplings. 

Importantly, increasing $C_{gj}$ lowers the frequencies of the lowest array modes, while simultaneously enhancing the coupling strengths $g_{f\mu}$ and $g_{r\mu}$. In practice, the first two array modes ($\mu = 1,2$) are the closest in frequency to both the resonator and the qubit, and exhibit stronger coupling than higher modes of the same parity. Accordingly, we restrict the analysis to these two modes, neglecting modes with $\mu > 2$. Furthermore, by parity considerations, the second array mode couples more strongly to the qubit than the first. For our choice of parameters, the first (second) array mode frequency ranges between $\omega_{\mu = 1} / 2 \pi \simeq 20$ GHz ($\omega_{\mu = 2} / 2 \pi \simeq 22$ GHz) and $\omega_{\mu = 1} / 2 \pi \simeq 6.3$ GHz ($\omega_{\mu = 2} / 2 \pi \simeq 8.2$ GHz) as the ground capacitance is increased from  $C_{gj} = 0.01$ fF to $0.5$ fF. Further details on the array mode parameters are discussed in \cref{app:circuit_params}.

With the array modes accounted for, the total Hilbert space is larger, and we therefore rely on a Floquet branch analysis to identify the critical photon numbers. Following Ref.~\cite{Dumas2024}, the resonator and its drive are replaced by an effective drive directly on the fluxonium; see~\cref{app:Floquet BA} for details. In the Floquet framework, the drive frequency plays the role of the resonator frequency in the branch analysis. In this case, the analysis labels and tracks the Floquet modes of the driven fluxonium and of the array modes. As a result, we also update our definition for the critical photon number as the drive amplitude at which the ground (excited) state of the fluxonium exceeds $\langle \h{N}_f \rangle = 2$ ($3$) \textit{or} when any of the array mode's population exceeds $\langle \h{N}_{a, \mu} \rangle = 0.3$. The choice of a lower threshold for the array mode is to avoid strong dressing with the array modes. Indeed, populating the array modes leads to dephasing of the qubit~\cite{singh:2025}.

\begin{figure*}[ht]
    \centering
    \includegraphics[width=\linewidth]{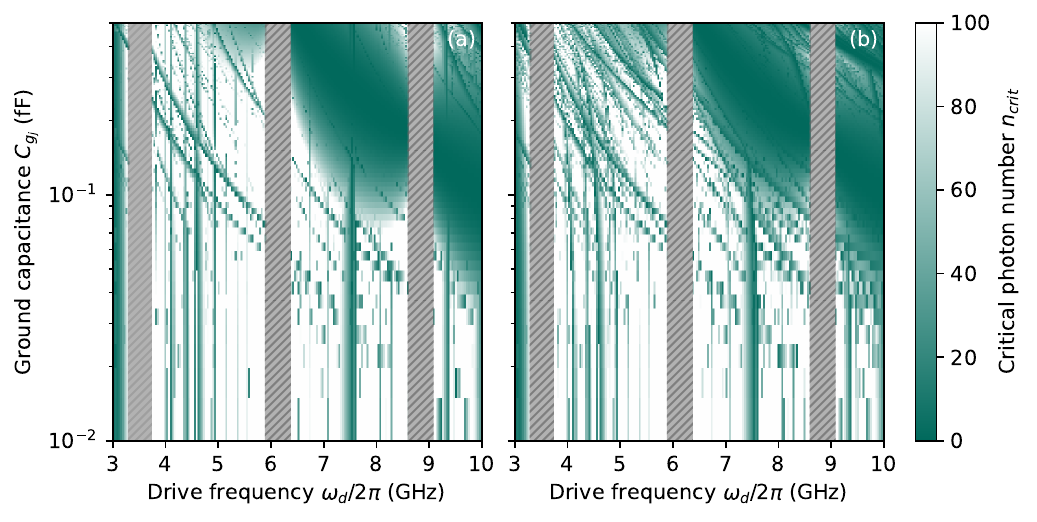}
    \caption{Critical photon numbers as a function of resonator (drive) frequencies $\omega_d$ and ground capacitance $C_{gj}$ with (a) the first $\mu = 1$ or (b) first and second $\mu = 2$ array modes included. The gray regions indicate where the dispersive approximation between the fluxonium and resonator are invalid. Here the fluxonium parameters are kept constant, and are $E_J / 2 \pi = 4.0$ GHz, $E_L / 2\pi = 0.75$ GHz, and $E_C / 2 \pi = 1.0$ GHz. 
    The coupling strengths $g_{fr}$ are chosen such that for all parameters the dispersive shift is fixed at $\chi/2\pi = 2.5$ MHz. 
    }
    \label{fig:array_modes}
\end{figure*}

\subsection{Critical photon numbers with array modes}

Following the approach outlined above, we extract the critical photon number as a function of the drive frequency and capacitance to ground $C_{gj}$ of the array islands which controls coupling to the array modes, see~\cref{fig:array_modes}. To see the effects of the array modes on a fluxonium qubit that has large average critical photon numbers in the absence of such modes, here we use our representative light fluxonium qubit, with $E_J / 2\pi, E_L / 2 \pi, E_C / 2 \pi = 4.0, 0.75, 1.0$ GHz. In panel (a) we account only for the first array mode while in panel (b) we account for both the first and second array modes. The gray shaded areas correspond to regions where the dispersive approximation fails. In both panels, the overall behavior is similar. The vertical resonances that emerge at small ground capacitances arise from multi-photon resonances that do not involve the array modes. In contrast, the diagonal bands of low critical photon numbers originate from multi-photon processes involving the array modes. We observe that these features appear at lower ground capacitance for drive frequencies closer to the array mode frequencies. Finally, the broad region of low critical photon numbers in the top-right corner of both panels arises due to the drive frequency approaching the array mode frequency.

One might expect the second array mode—because of its stronger coupling to the fluxonium—to generate more multi-photon resonances. However, our results indicate the opposite. Comparing~\cref{fig:array_modes}(a) and (b), we find that while the first array mode couples more weakly to the fluxonium, its lower frequency leads to more accidental multi-photon resonances. Although including the second array mode in~\cref{fig:array_modes}(b) makes the spectrum appear more cluttered, the overall qualitative behavior remains essentially unchanged. The large region of lower critical photon numbers in the top right of both panels indicates that the smaller detuning between the fluxonium and the first array mode makes it the dominant contributor to multiphoton processes.

The above results confirm that array modes play a central role in measurement-induced transitions of the fluxonium. An approach to mitigate their effects is to reduce the ground capacitance experienced by the Josephson junction array~\cite{singh:2025}. This not only pushes the array mode frequencies higher but also weakens their coupling to the fluxonium. Another approach is to increase the number of junctions $N$ while simultaneously scaling $E_{Jj}$ to keep $E_L$ fixed. This reduces the coupling between the array modes and the fluxonium, but lowers the frequencies of the first and second array modes. Conversely, decreasing the number of junctions increases the detuning between the fluxonium and the first and second array modes but simultaneously lowers the maximum array mode frequency, shifting all modes downward. In addition, reducing the junction count introduces nonlinear corrections to the formulas used in our analysis, adding further complexity~\cite{singh:2025, Viola:2015}. 

\begin{figure}
    \centering
    \includegraphics[width=\linewidth]{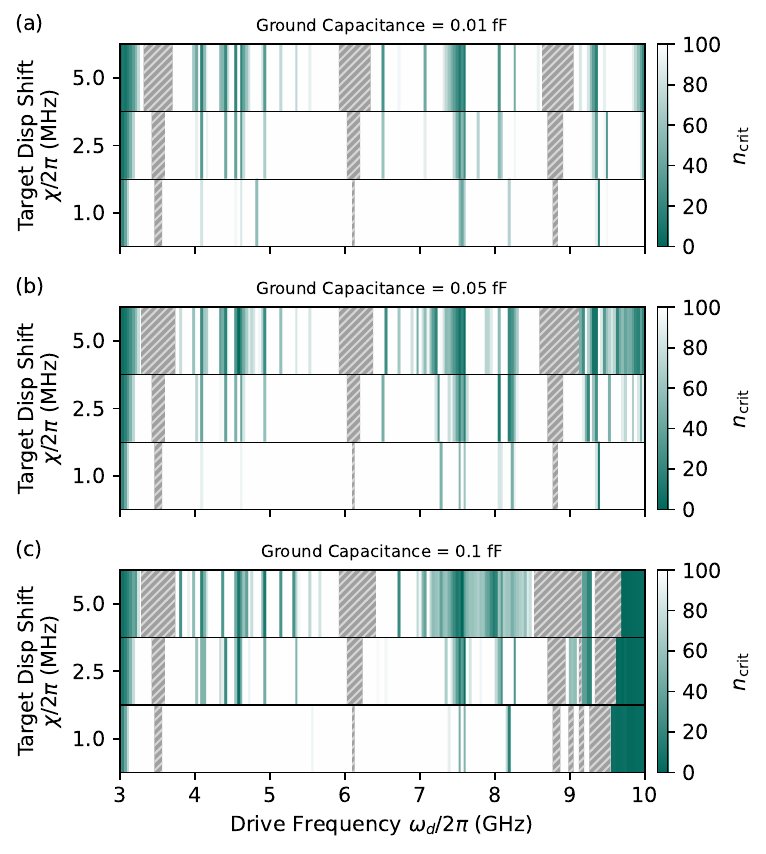}
    \caption{Critical photon numbers as a function of the resonator (drive) frequencies and target dispersive shifts $\chi$ for (a) $C_{gj} = 0.01$ fF, (b) $C_{gj} = 0.05$ fF, and (c) $C_{gj} = 0.1$ fF. The fluxonium parameters are the same as in \cref{fig:array_modes}. The gray regions indicate where the dispersive approximation between the fluxonium and resonator are invalid.
    }
    \label{fig:array_mode_disp_shift_compare}
\end{figure}

We have so far fixed the dispersive shift to be $\chi / 2\pi = 2.5$ MHz. However, given the uncertainty that can arise in the ground capacitance, this naturally raises the question whether accidental resonances can be largely avoided if the dispersive shift is lowered, which in turn lowers the coupling between the fluxonium and the array mode for a fixed ground capacitance value. Furthermore, this possibility is supported by the observation that the readout fidelity remains approximately unchanged as long as the product $\chi \bar n$ is kept constant. In~\cref{fig:array_mode_disp_shift_compare}, we compare the critical photon numbers as a function of the drive frequency, the dispersive shift and for three values of the ground capacitance. As expected, increasing the ground capacitance expands the regions of low critical photon numbers. Independently, increasing the dispersive shift $\chi$ also enlarges these regions. Notably, when the dispersive shift is reduced to $\chi / 2\pi = 1$ MHz, the influence of ground capacitance is minimal. This suggests that aiming for smaller dispersive shifts may help reduce the likelihood of resonances in situations where the ground capacitance—and thus the array modes—cannot be reliably controlled. In cases where the ground capacitances can be accurately predetermined at the design stage, our analysis may prove useful in avoiding low critical photon number regions in the spectrum. 

\section{Conclusion} \label{sec:conclusion} 

Using the branch analysis and the Floquet branch analysis, we have provided a systematic exploration of the physics of measurement-induced transitions in the fluxonium across a wide range of parameter space. Our results demonstrate that lighter fluxoniums are generally less prone to MIST than their heavier counterparts. We have identified three mechanisms explaining why this is the case: a smaller density of multi-photon resonances, a smaller average coupling required to reach the same dispersive coupling, and a more harmonic-oscillator like structure of the charge matrix elements suppresses paths to higher excited states. Using time-dependent numerical simulations of the readout, we have verified that the critical photon numbers we find from branch analysis are a good predictor of measurement-induced transitions. These simulations also suggest that high-fidelity readout for a well-chosen set of realistic parameters is possible, although residual leakage---not due to MIST-like processes---still limits the assignment error. Finally, we have also explored the impact of array modes for fluxoniums coupled asymmetrically to the readout resonator. There we found that, even though the first array mode is more weakly coupled than the second array mode, its lower frequency is what leads to a reduced critical photon numbers even at moderate values of the parasitic ground capacitance. Our findings can be used to guide the design of fluxoniums and associated readout circuitry.

To guide their practical use, the averaging over resonator frequency is introduced solely to enable a compact 2D representation of the critical photon number data, and the specific choice of a $700$ MHz bandwidth is not unique. For a given fluxonium $(E_J/E_C,E_L/E_C)$, the landscape of critical photon numbers as a function of $\omega_r$ and $g$ remains highly structured. The coarse-grained results presented here can thus help navigate this parameter space, but examining the underlying fine-grained structure remains important to ensure that specific parameter choices avoid deleterious multi-photon transitions.

\section{Acknowledgments}

The authors are grateful to Othmane Benhayoune Khadraoui, Chunyang Ding, Chuyao Tong, Paul Varosy, David Schuster, Danyang Chen, Tianpu Zhao, Sai Pavan Chitta, Jens Koch, Miguel Moreira, and Jorge Marques for insightful discussions. This research was sponsored by IARPA under the Entangled Logical Qubits program and funding from the U.S. Army Research Office Grant No. W911NF-23-1-0101. Additional support is acknowledged from NSERC, and the Ministère de l’Économie et de l’Innovation du Québec. 


\appendix

\section{Details on fluxonium parameter scan} \label{app_sec:scan_details}

In this section we outline how the critical photon numbers in \cref{fig:giga_scan} are computed. We first begin by choosing an appropriate range for the ratios $E_J/E_C$ and $E_L/E_C$. We chose $E_J/E_C$ to be between $1$ and $10$, and $E_L/E_C$ to be between $0.05$ and $3$. We take $10^2$ equally spaced values for both ratios, leading to $10^4$ different fluxonium qubits, as discussed in \cref{sec:extracting_ncrits}. These ranges were chosen to include typical parameters used in fluxonium experiments while still accounting for a wider range which could open up new avenues for fluxonium designs and yield a better understanding of the dependence of the critical photon number on the parameters.

The critical photon numbers heavily depend on the resonator frequency $\omega_r$. In our work, we restrict the resonator frequencies to be between the typical values $\omega_r / 2\pi = 3 \sim 10$ GHz. For each pair ($E_J/E_C$, $E_L/E_C$), we compute the critical photon numbers for $200$ equally spaced resonator frequencies. Thus, in total, we obtain the critical photon numbers from $2 \times 10^6$ branch analyses. 

Furthermore, as stated in \cref{sec:extracting_ncrits}, for each pair of ($E_J/E_C$, $E_L/E_C$, $\omega_r$) we pick the coupling strengths $g$ such that the dispersive shift is always $\chi / 2\pi = 2.5$ MHz. To obtain the  coupling strengths $g$, we thus first numerically diagonalize the system Hamiltonian for all permutations of ($E_J/E_C$, $E_L/E_C$, $\omega_r$), for $10^2$ coupling strengths ranging from $g/2\pi = 5$ to $500$ MHz. We then numerically extract the dispersive shift $\chi$ for each combination ($E_J/E_C$, $E_L/E_C$, $g$, $\omega_r$) and interpolate the results. Using the interpolated dispersive shifts, for each combination of ($E_J/E_C$, $E_L/E_C$, $\omega_r$), we estimate the coupling strength $g$ that gives the desired $\chi/2\pi = 2.5$ MHz. Note that although the coupling strengths are interpolated, this method can accurately estimate the dispersive shift to within a few $10$'s of kHz. Furthermore, in almost all cases the interpolation only fails to find an accurate estimate for the coupling strength when the parameters are such that the dispersive approximation is not valid, more on this below. Nonetheless, we filter out all points where the dispersive shift, calculated using the estimated coupling strength $g$, deviates from the target dispersive shift by more than $100$ kHz. This ensures that in all of our analyses, the dispersive shift is guaranteed to be within the range $\chi / 2\pi \in [2.4,2.6]$~MHz. 

To ensure that we are in the dispersive regime, for each combination of ($E_J/E_C$, $E_L/E_C$, $g$, $\omega_r$), we compute the dispersive approximation condition given in \cref{eq:dispersive_condition}, and reject all points that return a value greater than $0.15$. The value $0.15$ is chosen to strike a balance between ensuring that sufficient points in the spectrum are considered, and removing all points that lead to substantial hybridization between the fluxonium and the resonator at low photon numbers. 

The branch analyses were performed on a NVIDIA GH200 GPU (480GB). The fluxonium Hamiltonian is created in the Fock basis using $1000$ Fock states, and truncated to $20$ states after diagonalization. The resonator dimension was set to $200$, resulting in a Hilbert space size of $4000$. The system dimensions were chosen to balance convergence of results and simulation speed. In our hardware, each instance of the branch analysis took approximately 0.2 seconds to perform.

\section{Inductive coupling for fluxonium readout} \label{app_sec:inductive_coupling}

\begin{figure*}
    \centering
    \includegraphics[width=\linewidth]{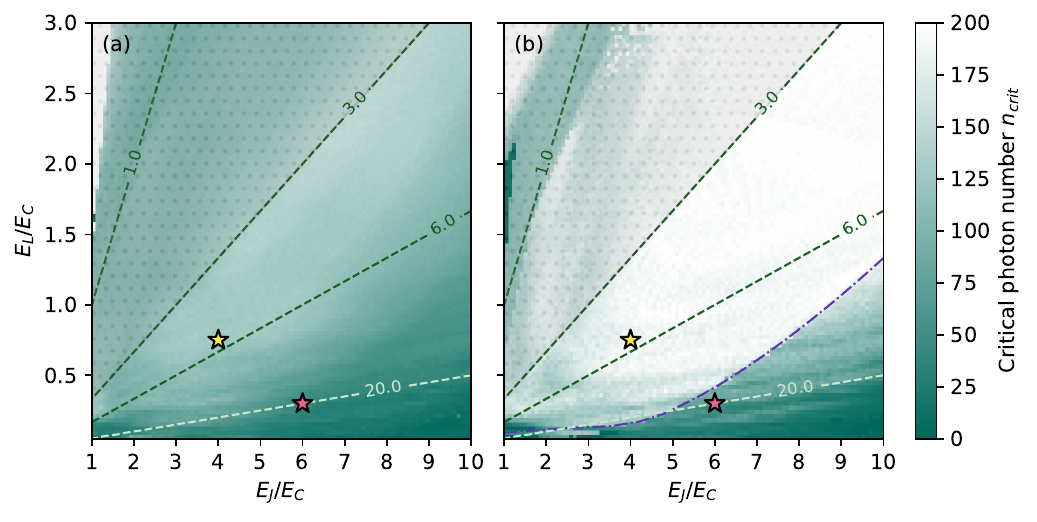}
    \caption{Same as \cref{fig:giga_scan} but for inductive coupling to the readout resonator.
    }
    \label{fig:gigascan_inductive}
\end{figure*}

Due to the large inductance of the fluxonium, inductive coupling to a readout resonator provides a natural coupling mechanism \cite{Gusenkova:2021, Rieger:2023}. In this section, we repeat for inductive coupling the analysis done in \cref{sec:fluxonium_parameter_scan} for capacitive coupling. Similarly to that analysis, we begin by finding the coupling strength $g$ leading to the target dispersive shift of $\chi/2\pi = 2.5$ MHz for each combination of ($E_J/E_C, E_L/E_C$, $\omega_r$). We then proceed to compute the critical photon numbers using branch analysis in the same manner. The average critical photon numbers over all resonator frequencies $\omega_r /2 \pi = [3, 10]$ GHz is shown in \cref{fig:gigascan_inductive}(a). Similarly, to \cref{fig:giga_scan}(b), \cref{fig:gigascan_inductive}(b) shows the average critical photon number across any $700$ MHz window of resonator frequencies for each combination of ($E_J/E_C, E_L/E_C$). 

In both \cref{fig:gigascan_inductive}(a) and (b), we observe features similar to those in \cref{fig:giga_scan}(a) and (b). In particular, the average critical photon number decreases as the $E_J/E_L$ ratio increases, for the same reasons discussed in the main text. We note that in general, inductive coupling tends to lead to slightly smaller critical photon numbers than capacitive coupling. These differences can arise, for example, because of differences in the matrix elements which can lead to slightly differing contributions to the dispersive shift from each level. Furthermore, in the heavy fluxonium regime the flux operator is more harmonic-oscillator like compared to the charge which may also contribute to differences in the critical photon numbers. Moreover, in almost all cases we find that $\vert \langle 0_f \vert \hat{\varphi}_f \vert 1_f \rangle \vert \gg \vert \langle 0_f \vert \hat{n}_f \vert 1_f \rangle \vert$. In fact, the flux $\vert \langle 0_f \vert \hat{\varphi}_f \vert 1_f \rangle \vert$ matrix element is often much greater in magnitude than $\vert \langle 1_f \vert \hat{\varphi}_f \vert 2_f \rangle \vert$ and $\vert \langle 0_f \vert \hat{\varphi}_f \vert 3_f \rangle \vert$. Especially in the case of heavy fluxoniums, the large $\vert \langle 0_f \vert \hat{\varphi}_f \vert 1_f \rangle \vert$ matrix element can lead to the breakdown of the dispersive approximation at resonator frequencies in the range $\omega_r / 2\pi \sim [3, 5]$ GHz. This leads to enhanced ac-Stark shifts, increasing the likelihood of frequency collisions and thereby reducing the critical photon number. In contrast, such a breakdown is not observed in the same frequency range for charge coupling, owing to the comparatively small $\vert \langle 0_f \vert \hat{n}_f \vert 1_f \rangle \vert$ matrix element.

Finally, in general fluxonium qubits have a more stringent capacitance budget compared to transmon qubits, due to their typically higher charging energy $E_C$. This capacitance budget can easily be exceeded when strong couplings to nearby qubits, couplers or the readout resonator is required. We note that, despite the somewhat reduced critical photon number, inductive coupling to the readout resonator can help relax constraints on the capacitance budget.

\section{Details on readout simulations} \label{app_sec:readout_simulations}

\begin{figure}
    \centering
    \includegraphics[width=\linewidth]{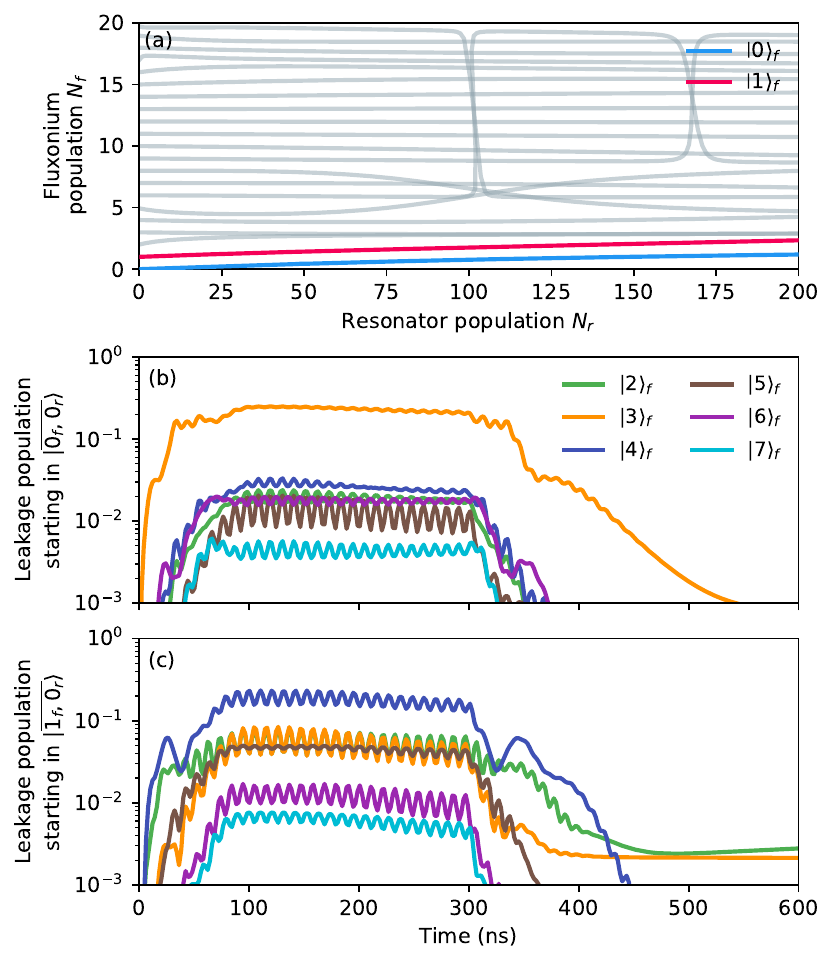}
    \caption{(a) Branch analysis for the parameters used in the readout simulations of \cref{fig:high_ncrit_readout_sim}. Leakage population when starting in (b) the ground or (c) the excited state. Here we show only the leakage to the first six states out of the computational subspace.}
    \label{fig:high_ncrit_readout_sim_detailed}
\end{figure}

In this section we present details of the time-dependent readout simulations discussed in \cref{sec:readout}. As mentioned in the main text, we solve the standard Lindblad master equation (c.f. \cref{eqn:master_eqn}) using the \textit{mesolve} solver from the \texttt{dynamiqs} package \cite{guilmin2025dynamiqs}. We employ a two-step pulse with a drive amplitude $\epsilon_1$ for the first $100$ ns and $\epsilon_2$ for the subsequent $200$ ns. We then let the resonator decay freely for $300$ ns. For the readout simulations shown in \cref{fig:readout_simulation}(b) and (c), we use the drive amplitudes $(\epsilon_1 / 2\pi, \epsilon_2 / 2\pi) = (24.5, 8), (40, 32.5)$ and $(65, 65)$ MHz with $\kappa / 2\pi = 2.5, 5$, and $10$ MHz, respectively. Those drive amplitudes were selected to bring the average resonator population to $\overline{n} = 40$ within the first $100$ ns, and to approximately sustain this population for the subsequent $200$ ns. In all cases, we keep $20$ states in the fluxonium, and $65$ for the resonator.

For the readout simulations of \cref{fig:high_ncrit_readout_sim} we keep the fluxonium qubit parameters the same as in \cref{fig:readout_simulation} but set the resonator decay rate to $\kappa / 2\pi = 5$ MHz. Furthermore, we use a two-step pulse with $\epsilon_1 / 2\pi= 55$ MHz and $\epsilon_2 / 2\pi = 47.5$ MHz. The steady-state population of the resonator reaches $\sim 70$ photons in $100$ ns. We keep $20$ states in the fluxonium, and $150$ states in the resonator. Finally, the resonator frequency is set to $\omega_r / 2 \pi = 7.925$ GHz and the coupling strength to $g/2\pi = 227.5$ MHz corresponding to a dispersive shift of $\chi/2\pi = 2.5$ MHz. The branch analysis for the parameters used in \cref{fig:high_ncrit_readout_sim} is shown in \cref{fig:high_ncrit_readout_sim_detailed}(a). As can be seen, there are no drive-induced transitions of the computational states for the full range that was computed, i.e., up to $200$ photons. Furthermore, \cref{fig:high_ncrit_readout_sim_detailed}(b) and (c) show the leakage population when starting in the ground or excited state. As mentioned in the main text, when starting in the ground state, most of the leakage goes to the third excited state. Conversely, when starting in the excited state, most of the leakage goes to the second and fourth excited state. 

Finally, we discuss the details of \cref{fig:high_ncrit_readout_sim}(b). Assuming that the resonator response is well described by a Gaussian in phase space, the assignment error can be calculated using the signal-to-noise ratio \cite{Bultink:2018} 
\begin{align}
    \mathrm{SNR}(t_\mathrm{m}) = \sqrt{2 \kappa \eta \int_0^{t_\mathrm{m}} \vert \langle \ha \rangle_e(t) - \langle \ha \rangle_g(t) \vert^2 \: dt},
\end{align}
where $\eta$ is the measurement efficiency and $t_m$ is the integration time. In this expression, $\langle \ha \rangle_{e/g}$ is the mean field trajectory of the resonator conditioned on the qubit starting in ground (g) or excited (e) state. The above formula assumes an optimal weight function is applied. The assignment error is then calculated as \cite{Gambetta:2007, Swiadek:2024}
\begin{align}
    \varepsilon(t_m) = \frac{1}{2} \mathrm{erfc}\left(\frac{\mathrm{SNR}(t_m)}{2 \sqrt{2}}\right) + \frac{t_m}{2 T_1},
\end{align}
where $T_1$ is the fluxonium's excited state lifetime. 

\section{Derivation of the array mode Hamiltonian}
\label{app_sec:array_modes}

In this appendix, we derive the Hamiltonian \cref{eq:H_f_array} corresponding to the circuit in \cref{fig:array_modes} describing the qubit, resonator, and array modes. Our derivation shares many similarities with those described in Refs.~\cite{Viola:2015, singh:2025, Ferguson:2013, Sorokanich:2024}.

\subsection{Phase drop coordinates}
Our starting point is the Lagrangian $\caL$ of the circuit given by 
\begin{align}
    \caL
    =
    \caL_{PS}
    +
    \caL_{JJA}
    +
    \caL_{g}
    +
    \caL_{r}
    +
    \caL_c,
\end{align}
where, using the node flux variables $\varphi_j$ coordinate system, we have
\begin{widetext}
\begin{align}
    \caL_{PS}
    &=
    \frac{1}{16E_{C_p}}
    (\dot{\varphi}_N - \dot{\varphi}_0)^2
    +
    E_{J_p}\cos(\varphi_N-\varphi_0 + \varphi_{\rm ext}),
    \\
    \caL_{JJA}
    &=
    \sum_{m=1}^N
    \frac{1}{16 E_{C_{g,j}}^{(m)}}
    (\dot{\varphi}_m - \dot{\varphi}_{m-1})^2
    +
    \sum_{m=1}^{N}
    E_{J_j}^{(m)}
    \cos(\varphi_m - \varphi_{m-1}),
    \\
    \caL_g
    &=
    \sum_{m=1}^{N-1}
    \frac{1}{16 E_{C_j}^{(m)}} \dot{\varphi}_m^2
    +
    \frac{1}{16 E_{C_{g,p}}}(\dot{\varphi}_0^2 + \dot{\varphi}_N^2),
    \\
    \caL_r
    &=
    \frac{1}{16 E_{C_r}}\dot{\varphi}_r^2 - \frac{E_{L_r}}{2}\varphi_r^2,
    \\
    \caL_c
    &=
    \frac{1}{16 E_{C_c}}(\dot{\varphi}_r - \dot{\varphi}_0)^2,
\end{align}
\end{widetext}
where $\caL_{PS}, \caL_{JJA}, \caL_{g}, \caL_{r}$ and $\caL_c$ are the Lagrangian of the phase slip junction, Josephson junction array, contributions from the ground, resonator, and coupling between the resonator and qubit respectively. The various charging energies are related to the capacitances in \cref{fig:array_modes} via the standard relation $E_{C}/2\pi = e^2/(2 C) $ \cite{Vool2017-gf}. 

\subsection{Cyclic variable}
The above expressions  account for fluxoid quantization by including the external flux $\varphi_{\rm ext}$ in the potential energy of the phase slip junction. There is also a non-dynamical degree of freedom in the coordinate system of choice, which is most obvious by writing the Lagrangian in the gauge-invariant phase drop coordinates $(\varphi_r, \varphi_0, \varphi_1, \dots, \varphi_N) \to (\varphi_r, \varphi_0, \theta_1, \dots, \theta_N)$ where, for $1 \leq m \leq N$ we have
\begin{align}
    \theta_m = \varphi_m - \varphi_{m-1}
\end{align}
which immediately implies
\begin{align}
    \varphi_m = \sum_{l=1}^m \theta_l + \varphi_0
\end{align}
for the same range $1 \leq m \leq N$. In this new coordinate system, the Lagrangian takes the form
\begin{widetext}
\begin{align}
    \caL_{PS}
    &=
    \frac{1}{16E_{C_p}}
    \left(
    \sum_{m=1}^N \dot{\theta}_m
    \right)^2
    +
    E_{J_p}\cos\left(\sum_{m=1}^N \dot{\theta}_m + \varphi_{\rm ext}\right),
    \\
    \caL_{JJA}
    &=
    \sum_{m=1}^N
    \frac{1}{16 E_{C_j}^{(m)}}
    \dot{\theta}_m^2
    +
    \sum_{m=1}^{N}
    E_{J_j}^{(m)}
    \cos \theta_m,
    \\
    \caL_g
    &=
    \sum_{m=1}^{N-1}
    \frac{1}{16 E_{C_j}^{(m)}} 
    \left( 
    \sum_{l = 1}^m
    \dot{\theta}_m
    +
    \dot{\varphi}_0
    \right)^2
    +
    \frac{1}{16 E_{C_{g,p}}}
    \left(
    \dot{\varphi}_0^2 
    + 
    \left( 
    \sum_{m=1}^N \dot{\theta}_m 
    +
    \dot{\varphi}_0
    \right)^2
    \right),
    \\
    \caL_r
    &=
    \frac{1}{16 E_{C_r}}\dot{\varphi}_r^2 - \frac{E_{L_r}}{2}\varphi_r^2,
    \\
    \caL_c
    &=
    \frac{1}{16 E_{C_c}}(\dot{\varphi}_r - \dot{\varphi}_0)^2.
\end{align}
\end{widetext}
With the coordinate $\varphi_0$ not appearing in the Lagrangian, the Euler-Lagrange equations immediately imply that $\partial \caL/\partial \dot{\varphi}_0 =0$. A simple calculation then leads to 
\begin{align}\label{app_eq:varphi_0}
    \dot{\varphi}_0
    =
    E_t
    \left(
    \frac{1}{E_{C_c}} \dot{\varphi}_r
    -
    \sum_{m=1}^N \frac{1}{E_{C,\theta_m}} \dot{\theta}_m
    \right),
\end{align}
where we have defined the total capacitive energy due to the ground and coupling capacitances as 
\begin{align}
    E_t
    =
    \left( 
    \sum_{m=1}^{N-1}
    \frac{1}{E_{C_{g,j}}^{(m)}}
    +
    \frac{2}{E_{C_{g,p}}}
    +
    \frac{1}{E_{C_c}}
    \right)^{-1},
\end{align}
as well as
\begin{align}
    E_{C,\theta_m}
    =
    \left( 
    \sum_{l = m}^{N-1} \frac{1}{E^{(l)}_{C_{h,j}}}
    +
    \frac{1}{E_{C_{g,p}}}
    \right)^{-1}.
\end{align}
With \cref{app_eq:varphi_0}, we can now obtain the Lagrangian as a function of the coordinates $\varphi_r$ and $\theta_n$ only.

Only $\caL_g$ and $\caL_c$ depend on $\dot{\varphi}_0$, while the remaining terms in the Lagrangian are unchanged. After another straightforward calculation, we can write 
\begin{align}
    \caL_g + \caL_c
    =
    \frac{1}{2}
    \sum_{l,l' = 1}^N
    \mathcal{G}_{ll'} \dot{\theta}_l \dot{\theta}_{l'}
    +
    \sum_{l = 1}^N \mathcal{F}_l \dot{\theta}_l \dot{\varphi}_r
    +
    \frac{1}{16 E_{C_c}} \dot{\varphi}_r^2
\end{align}
where we have defined  
\begin{align}\label{app_eq:G_ll}
    \mathcal{G}_{ll'}
    &=
    \frac{1}{8 E_{C, \theta_{{\rm{max}}(l,l')}}}
    \left(1- \frac{E_t}{E_{C, \theta_{{\rm{\min}}(l,l')}}} \right),
    \\ \label{app_eq:F_l}
    \mathcal{F}_{l}
    &=
    \frac{E_t}{8 E_{C, \theta_l, E_{C_c}}}.
\end{align}
Note that although we have dealt with a generic array so far, we now make the simplifying assumption that all junctions are the same. Using this assumption, we then have 
\begin{align}
    E_{t}
    &=
    \left( 
    \frac{N-1}{E_{C_{g,j}}}
    +
    \frac{2}{E_{C_{g,p}}}
    +
    \frac{1}{E_{C_c}}
    \right)^{-1},
    \\
    E_{C, \theta_m}
    &=
    \left( 
    \frac{N-m}{E_{C_{g,j}}} + \frac{1}{E_{C_{g,p}}}
    \right)^{-1},
    \\
    \mathcal{G}_{mm'}
    &=
    \frac{E_t}{8 E_{C, \theta_{{\rm max}(m,m')}}}
    \left( 
    \frac{1}{E_{C, \theta_{N+1- {\rm{min}}(m,m')}}}
    +
    \frac{1}{E_{C_c}}
    \right).
\end{align}
As expected from \cref{fig:array_modes}, the coupling capacitance breaks the inversion symmetry of the capacitance matrix $\mathcal{G}_{ll
'} \neq \mathcal{G}_{N+1-l, N+1-l'}$. 

The kinetic $\mathcal{T}$ and potential $\mathcal{U}$ energies are 
\begin{widetext}
    \begin{align}\label{app_eq:T}
        \mathcal{T}
        &=
        \frac{1}{16 E_{C_p}} 
        \left( \sum_{m=1}^N \dot{\theta}_m \right)^2
        +
        \frac{1}{16 E_{C_j}}
        \sum_{m=1}^N \dot{\theta}_m^2
        +
        \frac{1}{2}
        \sum_{m,m' = 1}^N
        \mathcal{G}_{mm'} \dot{\theta}_m \dot{\theta}_{m'}
        +
    \sum_{l = 1}^N \mathcal{F}_l \dot{\theta}_l \dot{\varphi}_r
    +
    \frac{1}{16 E_{C_c}} \dot{\varphi}_r^2,
    \\
    \mathcal{U}
    &=
    -E_{J_p}
    \cos \left( 
    \sum_{m=1}^N \theta_m
    +
    \varphi_{\rm ext}
    \right)
    -
    E_{J_j}\sum_{m=1}^N \cos \theta_m
    +
    \frac{E_{L_r}}{2}\varphi^2_r
    \end{align}.
\end{widetext}
from which we can then write the Lagrangian as the difference of the two $\caL = \mathcal{T} - \mathcal{U}$.

\subsection{Qubit and array modes}
We now want to write the Lagrangian in the qubit and array modes coordinate system. To that end, we define 
\begin{align}\label{app_eq:theta_m_new_basis}
    \theta_m
    =
    \frac{1}{N}
    \varphi_f
    +
    \sum_{\mu = 1}^{N-1}
    W_{\mu m} \xi_\mu
\end{align}
where $W_{\mu m}$ is an $(N-1) \times N$ matrix which satisfies
\begin{align}\label{app_eq:W_orthonormal}
    \sum_{m = 1}^N W_{\mu m} W_{\nu m} &= \delta_{\mu \nu},
    \\ \label{app_eq:W_phi_orthogonal}
    \sum_{m = 1}^N W_{\mu m} &= 0.
\end{align}
The first condition ensures the array modes are orthonormal, and the second ensures they are orthogonal to the qubit mode. Using these properties, we can write
\begin{align}
    \varphi_f &= \sum_{m=1}^N \theta_m,
    \\
    \xi_\mu &= \sum_{m=1}^NW_{\mu m} \theta_m,
\end{align}
where $\varphi_f$ will serve as the qubit degree of freedom and the $\xi_\mu$ are the array mode degrees of freedom which are entirely specified by the matrix $W_{\mu m}$. Throughout, we will use a convention where Roman indices run from 1 to $N$ and Greek indices run from 1 to $N-1$.

To obtain the Hamiltonian of the system, we perform a Legendre transformation on the Lagrangian. It is thus convenient to first write the kinetic energy term using matrix notation. To that end, we introduce the unnormalized fluxonium mode column vector
\begin{align}
    |\varphi_f)
    =
    \begin{pmatrix}
        1 \\ \vdots \\ 1
    \end{pmatrix}
\end{align}
as well as the array mode  column vectors 
\begin{align}
    |\xi_\mu)
    =
    \begin{pmatrix}
        W_{\mu1} \\
        \vdots \\
        W_{\mu N}
    \end{pmatrix},
\end{align}
which, via the properties \cref{app_eq:W_orthonormal} and \cref{app_eq:W_phi_orthogonal} satisfy $(\xi_\mu| \xi_\nu ) = \delta_{\mu \nu}$ and $(\phi | \xi_\mu) = 0$. It is also useful to note that the matrix $\boldsymbol{W}$ can be written as
\begin{align}
    \boldsymbol{W} = 
    \begin{pmatrix}
        (\xi_1|
        \\
        \vdots
        \\
        (\xi_{N-1}|
    \end{pmatrix}.
\end{align}
This then allows us to succinctly express the phase-drop coordinate system to the qubit and array mode coordinate system
\begin{align}\label{app_eq:coordinate_change_matrix}
     \dot{\vec{\theta}}
    =
    \left(
     \frac{1}{N}|\phi) \: \boldsymbol{W}^T
    \right)
    \begin{pmatrix}
       \varphi_f \\ \vec{\xi}
    \end{pmatrix},
\end{align}
where $\vec{\theta} = (\theta_1, \dots, \theta_N)^T$ is column vectors of the phase drop coordinates and $\vec{\xi} = (\xi_1, \dots, \xi_{N-1})^T$ is a column vector of array modes.

With this notation in hand, we can write the kinetic energy as
\begin{align}
    \mathcal{T}
    =
    \frac{1}{2}
    \begin{pmatrix}
        \dot{\varphi}_r \:\dot{\vec{\theta}}^T
    \end{pmatrix}
    \boldsymbol{C}
    \begin{pmatrix}
    \dot{\varphi}_r \ \\ \dot{\vec{\theta}}
    \end{pmatrix},
\end{align}
where and $\boldsymbol{C}$ is the capacitance matrix in the phase drop coordinate system. It takes the form
\begin{align}
    \boldsymbol{C}
    =
    \begin{pmatrix}
        C_{\varphi_r \varphi_r} & (\mathcal{F}|\\
        |\mathcal{F}) & \boldsymbol{C}_{\theta \theta}
    \end{pmatrix},
\end{align}
where $C_{\varphi_r \varphi_r} = (8 E_{C_r})^{-1} + (8 E_{C_c})^{-1}$, $|\mathcal{F}) = (\mathcal{F}_1, \dots, \mathcal{F}_N)^T$ is the column vector of coefficients defined in \cref{app_eq:F_l}, and 
\begin{align}
    \boldsymbol{C}_{\theta \theta}
    =
    \frac{1}{8 E_{C_p}} |\varphi_f) (\varphi_f|
    +
    \frac{1}{8 E_{C_j}} \boldsymbol{1}
    +
    \boldsymbol{\mathcal{G}},
\end{align}
where $\boldsymbol{1}$ is the $ N \times N$ identity matrix and $\boldsymbol{\mathcal{G}}$ is the matrix whose coefficients is defined in \cref{app_eq:G_ll}.

In the qubit and array mode coordinate system, we can use \cref{app_eq:coordinate_change_matrix} to write the kinetic energy as
\begin{align}
    \mathcal{T}
    =
    \frac{1}{2}
    \begin{pmatrix}
        \dot{\varphi}_r \: \dot{\varphi}_f \: \dot{\vec{\xi}}^T
    \end{pmatrix}
    \boldsymbol{C}'
    \begin{pmatrix}
        \dot{\varphi}_r \\ \dot{\varphi}_f \\ \dot{\vec{\xi}}^T
    \end{pmatrix}
\end{align}
where $\boldsymbol{C}'$ is the capacitance matrix in this new coordinate system and can be written as
\begin{align}
    \boldsymbol{C}'
    =
    \begin{pmatrix}
        C_{\varphi_r \varphi_r} & \frac{1}{N}(\mathcal{F}|\varphi_f) & [\boldsymbol{W} |\mathcal{F})]^T   \\ 
        \frac{1}{N}(\varphi_f|\mathcal{F}) & \frac{1}{N^2}(\varphi_f | \boldsymbol{C}_{\theta \theta} |\varphi_f) & \frac{1}{N}[\boldsymbol{W} \boldsymbol{C}_{\theta \theta} |\varphi_f) ]^T \\
        \boldsymbol{W}|\mathcal{F}) & \frac{1}{N}\boldsymbol{W} \boldsymbol{C}_{\theta \theta} |\varphi_f) & \boldsymbol{W} \boldsymbol{C}_{\theta \theta} \boldsymbol{W}^T
    \end{pmatrix}.
\end{align}
Note that the elements of the various vectors and matrices in $\boldsymbol{C}'$ can be expressed using overlaps, e.g., the elements of $\boldsymbol{W} \boldsymbol{C}_{\theta \theta} \boldsymbol{W}^T$ and be written as $(\xi_\mu| \boldsymbol{C}_{\theta \theta} |\xi_\nu)$.

To compute this quantity explicitly, we thus need to choose a set of array modes, which as previously mentioned, is equivalent to fixing a matrix $\boldsymbol{W}$. Given the translational invariance in the bulk of the array, it useful to choose the array modes which are also translationally invariant in the bulk. The choice of standing waves
\begin{align}\label{app_eq:W}
 W_{\mu m}
 =
 \sqrt{\frac{2}{N}}
 \cos 
 \frac{\pi \mu (m-\frac{1}{2})}{N}
\end{align}
is standard in the literature \cite{Ferguson:2013}. Recently, it was shown in Ref~\cite{Sorokanich:2024} that this choice is indeed well-motivated: the eigenvectors of the capacitance matrix of the array (the lower-right $N\times N$ matrix of $\boldsymbol{C}$) are in fact standing waves, except their wavelength depends on the parameters of the array. The strategy is then to express $\boldsymbol{C}'$ with this choice of coordinates, and then use perturbation theory to compute the inverse of the capacitance matrix.

\subsection{Perturbation theory}
Given the choice of array modes, we can compute the matrix elements of $\boldsymbol{C}'$ analytically. We have
\begin{widetext}
\begin{align}
    C_{\varphi_r \varphi_r}
    &=
    \frac{1}{8 E_{C_r}}
    +
    \frac{1}{8 E_{C_c}},
    \\
    \frac{1}{N} (\mathcal{F}|\varphi_f)
    &=
    \frac{1}{16 E_{C_c}} 
    \left(
    1
    -
    \frac{E_t}{E_{C_c}}
    \right),
    \\
    (\xi_\mu| \mathcal{F}) 
    &=
    \frac{E_t}{8 E_{C_c} E_{C_{g,j}} \sqrt{2 N}}
    \frac{o_\mu c_\mu}{s_{\mu}^2},
    \\
    \frac{1}{N^2}(\varphi_f|\boldsymbol{C}_{\theta \theta} | \varphi_f)
    &=
    \frac{1}{8 E_{C_p}}
    +
    \frac{1}{8 N E_{C_j}}
    +
    \frac{1}{32 E_t}
    \left( 
    1 
    -
    \frac{2}{3}
    \frac{E_t(N-1)(N+1)}{E_{C_{g,j}}N}
    -
    \frac{E_t^2}{E^2_{C_c}}
    \right),
    \\
    \frac{1}{N}(\xi_\mu | \boldsymbol{C}_{\theta \theta} |\varphi_f)
    &=
    - \frac{1}{16 E_{C_g,j}}
    \frac{c_\mu}{\sqrt{2 N} s_\mu^2}
    \left( 
    o_{\mu+1} - o_{\mu} \frac{E_t}{E_{C_c}}
    \right),
    \\
    (\xi_\mu | \boldsymbol{C}_{\theta \theta} | \xi_\nu )
    &=
    \frac{1}{8 E_{C_j}} \delta_{\mu \nu}
    +
    \frac{1}{32 E_{C_{g,j}} s_{\mu} s_{\nu}}
    \left( 
    \delta_{\mu \nu}
    -
    \frac{2 E_t}{E_{C_{g,j}}N}
    \frac{o_\mu o_\nu c_\mu c_\nu}{s_\mu s_\nu}
    \right)
\end{align}
\end{widetext}
where $o_\mu = (1-(-1)^\mu)/2$, $c_{\mu} = \cos(\pi \mu/(2N))$, and $s_\mu = \sin(\pi \mu /(2 N))$.

The classical Hamiltonian of the system is obtained after performing a Legendre transformation on the Lagrangian. Given that the Lagrangian is quadratic in the time derivatives of the coordinates, the kinetic part of the Hamiltonian can generically be written as  
\begin{align}\nonumber
    H_{kin}
    &=
    4 E_{C_f} n_f^2
    +
    4 E_{C_r} n_r^2
    +
    4 \sum_{\mu}
    E_{C_\mu}n_{\mu}^2
    \\ \nonumber 
    &+
    J_{r f}n_r n_f
    +
    \sum_{\mu}
    J_{r \mu} n_{r} n_{\mu}
    +
    \sum_{\mu}
    J_{f \mu} n_{f} n_{\mu}
    \\
    &+
    \sum_{\mu < \nu}
    J_{\mu \nu} n_{\mu} n_{\nu}
    \\
    &=
    \frac{1}{2}
    \vec{n}^T
    \boldsymbol{C}'^{-1}
    \vec{n}.
\end{align}
Here, $\vec{n} = (n_r, n_f, n_{\mu =1}, \dots, n_{\mu = N-1})^T$ is a column vector and $n_{f}$, $n_r$, and $n_\mu$  are the canonically conjugate momenta to the qubit, resonator, and array modes, respectively. The first line thus corresponds to the charging energy of the qubit, resonator, and array modes, respectively. The last two are couplings between all of these different degrees of freedom. The charging energies and coupling coefficients can then be read off the inverse of the capacitance matrix. For instance, the diagonal components of $\boldsymbol{C'}^{-1}$ are eight times the charging energy.

Given that all off-diagonal components of $\boldsymbol{C}'$ are proportional to the inverse of the charging energy of the junctions to ground, which by assumption is large, we can perform perturbation theory to compute $\boldsymbol{C}^{-1}$. Using $(\boldsymbol{A} + \boldsymbol{B})^{-1} = \boldsymbol{A}^{-1} - \boldsymbol{A}^{-1} \boldsymbol{B} \boldsymbol{A}^{-1} + \dots$ for any pair of matrices $\boldsymbol{A}$ and $\boldsymbol{B}$, we set $\boldsymbol{A}$ as the diagonal component of $\boldsymbol{C}'$ and $\boldsymbol{B}$ as its off-diagonal component to obtain the charging energies to second order in the off-diagonal components of $\boldsymbol{C}'$ such that,
\begin{widetext}
\begin{align}
    E_{C_r}
    &=
    \frac{1}{8 C_{\varphi_r \varphi_r}}
    =
    \frac{E_{C_r} E_{C_c}}{E_{C_r} + E_{C_c}},
    \\
    E_{C_f}
    &=
    \frac{N^2}{8 (\varphi_f|\boldsymbol{C}_{\theta \theta} |\varphi_f)}
    =
    \left(
    \frac{1}{E_{C_p}}
    +
    \frac{1}{N E_{C_j}}
    +
    \frac{1}{4 E_t}
    \left( 
    1 
    -
    \frac{2}{3}
    \frac{E_t(N-1)(N+1)}{E_{C_{g,j}}N}
    -
    \frac{E_t^2}{E^2_{C_c}}
    \right)
    \right)^{-1}, \label{app_eq:Ecf}
    \\
    E_{C_\mu}
    &=
    \frac{1}{8 (\xi_\mu | \boldsymbol{C}_{\theta \theta} | \xi_\mu )}
    =
    \left(
    \frac{1}{E_{C_j}}
    +
    \frac{1}{4 E_{C_{g,j}} s_{\mu}^2}
    \left( 
    1
    -
    \frac{2 E_t}{E_{C_{g,j}}N}
    \frac{o_\mu c_\mu^2}{s_\mu^2}
    \right)
    \right)^{-1},
\end{align}
with the couplings coefficients
\begin{align}
    J_{rf}
    &=
    -64E_{C_r} E_{C_f}
    \frac{1}{N} (\mathcal{F} | \varphi_f)
    =
    -\frac{4E_{C_r} E_{C_f}}{ E_{C_c}}
    \left(
    1- \frac{E_t}{E_{C_c}}
    \right), \label{app_eq:Jrf}
    \\
    J_{r\mu}
    &=
    -64E_{C_r} E_{C_\mu} (\mathcal{F}| \xi_\mu )
    =
    - \frac{8 E_{C_r} E_{C_\mu} E_t}{E_{C_c} E_{C_{g,j}}\sqrt{2 N}} \frac{o_\mu c_\mu}{s_\mu^2},
    \\
    J_{f \mu}
    &=
    -64 E_{C_f}E_{C_\mu} \frac{1}{N} (\varphi_f| \boldsymbol{C}_{\theta \theta} |\xi_\mu)
    =
    \frac{4 E_{C_f} E_{C_\mu}}{E_{C_{g,j}}} \frac{c_\mu}{\sqrt{2 N} s_\mu^2}
    \left(o_{\mu+1} - o_\mu \frac{E_t}{E_{C_c}} \right),
    \\
    J_{\mu \nu}
    &=
    -64 E_{C_\mu}E_{C_\nu}
    (\xi_\mu | \boldsymbol{C}_{\theta \theta} | \xi_\nu)
    =
    \frac{4 E_{C_\mu}E_{C_\nu} E_t}{E_{C_{g,j}}^2 N}
    \frac{o_\mu o_\nu c_{\mu}c_{\nu}}{s_{\mu}^2 s_{\nu}^2}.
\end{align}
\end{widetext}
Note that in the main text we ignore the coupling $J_{\mu \nu}$ between the array modes. This term only leads to a renormalization of frequency and mode structure of the array, which will come to quadratic order in the small off-diagonal matrix elements. To compare with the results in the main text, we also have to include the appropriate zero-point fluctuations of the charge operators. For the resonator, this is given by $n_{{\rm zpf}, r} = (E_{C_r}/(32 E_{L_r}))^{1/4}$, whereas $n_{{\rm zpf}, \mu} = (E_{C_\mu}/(32 E_{J_j}))^{1/4}$ where we have made the standard approximation and treated the array modes as linear. Comparing with \cref{eq:H_f_array} we have
\begin{align}
    \omega_\mu
    &=
    \sqrt{8 E_{C_\mu} E_{J_j}},
    \\
    g_{rf} 
    &=
    J_{rf} n_{{\rm zpf}, r}, \label{app_eq:grf}
    \\
    g_{r\mu}
    &=
    J_{r\mu} n_{{\rm zpf}, r} n_{{\rm zpf}, \mu},
    \\
    g_{f\mu}
    &
    =
    J_{f\mu} n_{{\rm zpf}, \mu}.
\end{align}

\subsection{Circuit parameters} \label{app:circuit_params}

The circuit parameters used in \cref{sec:array_modes} are summarized in \cref{table:circuit_params}. As can be seen in \cref{app_eq:Ecf}, the total charging energy of the fluxonium depends on the array junctions' capacitance to ground $C_{gj}$, among other circuit parameters. We choose these to ensure that the fluxonium's charging energy is fixed at $E_{Cf} / 2\pi = 1$ GHz for all values of $C_{gj}$.

Specifically, for each drive frequency $\omega_d$, we first find the coupling strength $g_{fr}$ that results in a dispersive shift $\chi/2\pi = 2.5$ MHz as was discussed in \cref{sec:extracting_ncrits}. Then, for each combination $(\omega_d, g_{fr}, C_{gj})$, we can use \cref{app_eq:grf}, \cref{app_eq:Jrf}, and \cref{app_eq:Ecf} to compute $E_{Cc}$ and $E_t$. From there, all other quantities such as the array mode frequencies and their coupling strengths to the fluxonium can be computed. That is, by enforcing $E_{Cf} / 2\pi = 1$ GHz, and a predetermined coupling strength $g_{rf}$, the capacitances $C_p$ and $C_c$ can be determined.

\begin{figure}
    \centering
    \includegraphics[width=\linewidth]{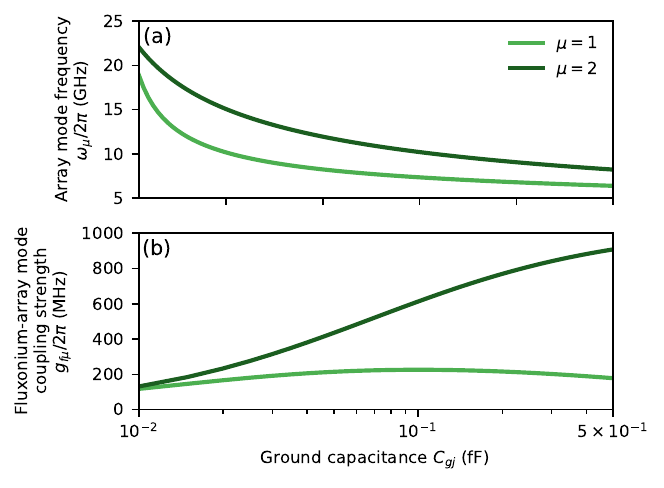}
    \caption{(a) Array mode frequencies and (b) their coupling strengths to the fluxonium. Here, the resonator frequency is set to $\omega_d / 2\pi = 7$ GHz, and the fluxonium–resonator coupling strength is $J_{rf} / 2\pi = 57$ MHz.}
    \label{fig:array_mode_freqs}
\end{figure}

Finally, \cref{fig:array_mode_freqs}(a) and (b) display the frequencies $\omega_\mu$ of the first and second array modes, along with their corresponding coupling strengths $g_{f\mu}$ to the fluxonium. As discussed in \cref{sec:array_modes}, the first array mode has a lower frequency and weaker coupling to the fluxonium, whereas the second mode exhibits a higher frequency but stronger coupling to the fluxonium.

\begin{table}[]
\begin{tabular}{|c|c|c|}
\hline
\textbf{Parameters} & \textbf{Variables} & \textbf{Values} \\ \hline
Array Josephson energy          & $E_{Jj} / 2\pi$ & $90$ GHz \\ \hline
Junction array count            & $N$ & $120$ \\ \hline
Array junction capacitance      & $C_j$ &  $25$ fF  \\ \hline
Phase slip junction capacitance &  $C_{gp}$  &  $10$ fF  \\ \hline
Resonator impedance             & $Z_r$      &  $50 \: \Omega$ \\ \hline
\end{tabular}
\caption{Summary of circuit parameters used in \cref{sec:array_modes}.}
\label{table:circuit_params}
\end{table}

\section{Floquet branch analysis}
\label{app:Floquet BA}

In this section we briefly describe the tool of Floquet branch analysis \cite{Cohen2023, Dumas2024, Xiao2024}. Our starting point is the driven version \cref{eq:H_f_array} 
\begin{align}
    \hat{H}
    &=
    \omega_r \hat{a}^\dagger \hat{a}
    +
    \hat{H}_f
    -i g_{rf}(\ha-\ha^\dagger) \hat{n}_f
    +
    \hat{H}_{arr}
    +
    \hat{H}_{int} \nonumber \\ 
    &-
    i \epsilon(t)(\ha - \had)\cos{\omega_d t},
\end{align}
where $\omega_d$ is the drive frequency and $\epsilon(t)$ is the time-dependent drive amplitude. 

The drive on the resonator leaves it approximately in a coherent state with amplitude $\alpha(t)$, where $\alpha(t)$ is determined by the linear response of the resonator \cite{Cohen2023}. Performing a displacement transformation on the resonator $\hat{a} \to \hat{a} + \alpha(t)$ and ignoring the backaction and all fluctuations, we drop the resonator entirely; the entire effect of the resonator is approximated by a charge drive on the fluxonium and array modes. Assuming $\omega_d \approx \omega_r$, the amplitude $\alpha(t)$ factorizes into a fast oscillating part with frequency $\omega_d$ and a slowly-varying envelope. Taking this envelope to be constant over the period of the drive, we arrive at a semiclassical Hamiltonian of the form \cite{Dumas2024, singh:2025}

\begin{align}
    \hat{H} &= \hat{H}_f + H_{arr} -i \sum_\mu g_{f\mu} \hat{n}_f (c_\mu - c_\mu^\dagger) \nonumber \\
    &- 2 g_{rf} \sqrt{\overline{n}_r} \hat{n}_f \cos{\omega_d t} \nonumber \\
    &+ 2 i \sum_\mu g_{r\mu} \sqrt{\overline{n}_r} (c_\mu - c_\mu^\dagger) \cos{\omega_d t}.
    \label{eqn:semiclassical_hamiltonian}
\end{align}
For a detailed discussion of this transformation, we refer the reader to Ref.~\cite{Dumas2024}. In the above semiclassical Hamiltonian, the first line describes the fluxonium and array mode's interaction, which preserves a fully quantum mechanical description. In the last two lines, the resonator has been replaced with an effective classical drive on both the fluxonium and array modes, each proportional to their coupling to the resonator as well as the average number of photons in the resonator $\sqrt{\overline{n}_r}$.  

To perform the Floquet branch analysis, we first diagonalize the fluxonium–array mode subsystem and label the states as $\ket{\overline{i_f, j_\mu, k_\mu}}_{l=0}$ using the quantum branch analysis procedure outlined in \cref{sec:Recap_ionization}. Here, $j_\mu$ denotes the occupation of the first array mode, while $k_\mu$ denotes that of the second array mode. Following the same principle as in quantum branch analysis, we recursively track and label the eigenstates as the effective coupling strength is increased, or equivalently, as the average photon number $\bar{n}_r$ is increased. The subscript $l$ labels the iteration step of this procedure. In our simulations, $\bar{n}_r$ is incremented in steps of $0.5$ photons.

Given the periodically driven nature of the Hamiltonian, we compute and label the Floquet eigenstates of the system at each step. For a given increment of $\bar{n}_r$, we first obtain the Floquet eigenstates of \cref{eqn:semiclassical_hamiltonian}, denoted $\ket{\xi}_{l=1}$. For each initial state $\ket{\overline{i_f, j_\mu, k_\mu}}_{l=0}$, we identify the Floquet state $\ket{\xi}_{l=1}$ that maximizes the overlap with $\ket{\overline{i_f, j_\mu, k_\mu}}_{l=0}$ and label it as $\ket{\overline{i_f, j_\mu, k_\mu}}_{l=1}$. This procedure is then applied recursively as $\bar{n}_r$ is increased. In this way, we construct collections of Floquet states
\begin{align}
    B_{i_f, j_\mu, k_\mu} = \left\{ \ket{\overline{i_f, j_\mu, k_\mu}}_l \; \middle| \; \forall l \right\},
\end{align}
which we refer to as Floquet branches. Finally, in analogy with quantum branch analysis, we compute the average fluxonium and array mode populations along each branch as functions of $\bar{n}_r$, allowing us to extract the corresponding critical photon numbers.

\bibliography{articles}
\end{document}